\documentclass[11pt]{article}

\usepackage[english]{babel}
\usepackage[utf8x]{inputenc}
\usepackage{amsmath}
\usepackage{amsthm, amssymb}
\usepackage{amsfonts}
\usepackage{algorithm}
\usepackage{algorithmic}
\usepackage{graphicx}
\usepackage{color}
\usepackage{fullpage}
\usepackage{bbm}
\usepackage{tikz}
\usepackage{subcaption, url}
\usepackage[numbers]{natbib}

\newtheorem{theorem}{Theorem}

\newtheorem{definition}[theorem]{Definition}
\newtheorem{lemma}[theorem]{Lemma}
\newtheorem{corollary}[theorem]{Corollary}

\newtheorem*{mainquestion}{Main Question}

\newcommand{\revision}[1]{#1}
\newcommand{\revisiontwo}[1]{#1}

\newcommand{\abs}[1]{\left\vert{#1}\right\vert}

\newcommand{\lca}{\mathsf{lca}}

\DeclareMathOperator{\argmax}{argmax}

\newcommand{\R}{\mathbb{R}}                     % Reals.
\newcommand{\Z}{\mathbb{Z}}                     % Integers.
\newcommand{\M}{\mathcal{M}}                     % Matroids.
\renewcommand{\SS}{\mathbf{S}}

\newcommand{\G}{\ensuremath{\mathbf G}}
\newcommand{\E}{\ensuremath{\mathbf E}}

\newcommand{\MM}{\ensuremath{\mathbf M}}

\title{On the Construction of Substitutes}
\author{Eric Balkanski\\ Harvard University \and  Renato Paes Leme \\ Google
Research}
\date{}
\begin{document}

\maketitle

% note that the abstract must come before \maketitle
\begin{abstract}
 Gross substitutability is a central concept in Economics and is connected to
  important notions in Discrete Convex Analysis, Number Theory and the analysis
  of Greedy algorithms in Computer Science. Many different characterizations 
  are known for this class, but providing a constructive description remains
  a major open problem. The construction problem asks how to construct all gross
  substitutes from a class of simpler functions using a set of operations. Since
  gross substitutes are a natural generalization of matroids to real-valued
  functions, matroid rank functions form a desirable such class of simpler
  functions.
  
    Shioura proved that a rich class of gross
  substitutes can be expressed as sums of matroid rank functions, but it is
  open whether all gross substitutes can be constructed this way. Our
  main result is a negative answer showing that some gross substitutes cannot be
  expressed as positive linear combinations of matroid rank functions. En route,
  we provide necessary and sufficient conditions for the sum to preserve
  substitutability, uncover a new operation preserving substitutability and
  fully describe all substitutes with at most $4$ items.
\end{abstract}%

\maketitle
%%%%%%%%%%%%%%%%%%%%%%%%%%%%%%%%%%%%%%%%%%%%%%%%%%%%%%%%%%%%%%%%%%%%%%

\section{Introduction}

The concept of gross substitutes (GS) occupies a central place in different areas
such as economics, discrete mathematics, number theory and has been rediscovered
in many contexts. In Economics it was proposed by 
\citet{KelsoCrawford} as a sufficient (and in some sense
necessary \cite{GulStachetti}) condition for
Walrasian equilibrium to exist in economies with indivisible goods. The notion
also appears in the existence of stable matchings in two sided markets
\cite{HatfieldMilgrom, Roth84}, to design
combinatorial auctions \cite{ausubel2002ascending}, in the
study of trading networks \cite{hatfield2013stability,ikebe2015stability}, among
others. In fact, \citet{hatfield2015hidden} show that many
tractable classes of preferences with complementarities have a hidden
substitutability structure. The phenomenon of substitutes being embedded in more
complex settings is also present in \cite{sun2009double, ostrovsky2008stability}.

In discrete mathematics,  \citet{MurotaShioura} define the
concept of $M^\natural$-concave function which ports the concept of
convex functions from continuous domains to the discrete lattice, carrying over
various strong (Fenchel-type) duality properties (see \cite{murota2016discrete} for a recent survey on discrete convex analysis). In number theory,  \citet{DressWenzel} defined
the concept of valuated matroids to generalize the Grassmann-Pl\"{u}cker
relations in $p$-adic analysis. Those %turned out to 
correspond to the same
class of functions as shown by \citet{FujishigeYang}.

Finally, gross substitutes have a special role in computer science, since they
correspond to the class of set function $v:2^{[n]} \rightarrow \R$ for which the
optimization problem $\max_{S \subseteq [n]} v(S) - \sum_{i \in S} p_i$ can be
solved by the natural greedy algorithm for all $p \in \R^n$.

Given the central role that GS plays in  many different fields, understanding
its structure is an important problem. There have been many equivalent
characterizations of GS through time:
\citet{KelsoCrawford}, \citet{Murota96}, \citet{MurotaShioura}, \citet{DressTerhalle_WellLayered,
DressTerhalle_Rewarding}, \citet{ausubel2002ascending}, \citet{ReijnierseGellekomPotters},  \citet{LehmannLehmannNisan}, \citet{ben2017walrasian}. All the
previous characterizations define GS as the class of functions satisfying a
certain property. Yet,
providing a constructive description of GS remains an elusive open problem.
This is in sharp contrast with submodular functions, a more complex class in
many respects, but one that has a simple constructive description.

\paragraph{The construction problem.}
A constructive description consists of a base class of simpler functions
(e.g. unit demand or matroid rank functions) together with a set of operations
(e.g. sum, convolution, endowment, affine transformations) such that all GS
functions can be constructed from the base class by applying such operations.

The first version of this question was asked by  
\citet{HatfieldMilgrom}. They noted that most examples of substitutes arising in
practical applications could be described as valuations that are built from
assignment valuations (which are convolutions of unit-demand valuations) and the
endowment operation. They called this class \emph{Endowed Assignment
Valuations} (EAV)
and asked whether EAV exhaust all gross substitutes. 
\citet{ostrovskypaesleme2014} provided a negative answer to this question showing
that some matroid rank functions cannot be constructed using those operations.
They do so by adapting a result of \citet{brualdi1969comments}
on the structure of transversal matroids to the theory of gross substitutes. The
main insight in \cite{ostrovskypaesleme2014} is that unit-demand valuations are
not strong enough as a base class. They propose a class called \emph{Matroid
Based Valuations} (MBV) which are constructed from weighted matroid rank
functions. It is still unknown whether MBV exhausts the whole class of
substitutes or not.

Another important construction is due to  \citet{shioura2012matroid},
who provides a construction of a rich class
of valuations called \emph{Matroid Rank Sums}
(MRS) and shows that this class contains many
important examples of gross substitutes. Matroid rank sums are positive linear
combinations of matroid rank functions such that the matroids satisfy a \emph{strong quotient} property. In general positive linear combinations of
matroids are not GS but the strong quotient property provides a sufficient
condition for this to be true.

\paragraph{GS and matroids.}
Gross substitutes are similar to matroids in many respects.
%, three of which we discuss next and formalize in Section~\ref{}.
First, matroids can be
described as the subset systems that can be optimized via greedy algorithms
while GS is the collection of real-valued set functions that can be optimized using these same greedy algorithms. Another similarity
is that both classes can be described via the exchange property. Finally, when we restrict our attention to GS functions where
marginals are in $\{0,1\}$ we obtain exactly the set of matroid rank functions.
These similarities, among others, explain why GS are seen as the natural extension of matroids to
real-valued functions.

The lack of constructive characterization for GS, combined with GS naturally generalizing matroids, begs the following question.

\begin{center}
\emph{Can gross substitutes be constructed from matroids?}
\end{center}

 In other words, do all GS functions look like matroids or does the lift
from subset systems to real-valued functions produce significantly different
functions? Similarly as for Shioura's construction, we focus on positive
linear combinations and ask the question of whether all gross substitutes are
positive linear combinations of matroids.

\paragraph{Our results.} Our main result is a negative answer to the question
above. We show that not all gross substitute functions can be constructed via
positive linear combinations from matroid rank functions. This implies in
particular that MRS does not exhaust the class of GS functions.

The proof consists of exhibiting a GS valuation function over $5$ elements
that cannot be expressed as a positive linear combination of matroids. We note that
there are $406$ matroids over $5$ elements ($38$ up to isomorphism) and that a
GS function over $5$ elements is defined by $40$ non-linear conditions.
The search space is huge, continuous and non-convex, so 
solving it by enumeration is infeasible even for $5$ elements.

Our techniques involve building a combinatorial and polyhedral understanding of
GS functions. In fact, we prove that for $3$ and $4$ elements, all GS
functions can be written as convex combinations of matroids. In this process,
we provide a polyhedral understanding of the set of GS functions and give
necessary and sufficient conditions for the positive combination of two GS
functions to be GS. We obtain the counter-example for $n=5$ by carefully
understanding where the techniques used for proving the $n=3$ and $4$ fail.

Upon obtaining an example for which the proof technique fails, we still need to
argue that it is not in the convex combination of matroids. One way to do it is
to enumerate over the $406$ matroids over $5$ elements and solve a large linear
program. That would be a valid approach, but one that would make verifying
correctness a much more complicated task. Instead, we use a combination of
linear algebra and combinatorial facts about matroids to provide a complete
mathematical proof that our counter-example is indeed not a convex combination
of matroids.

\paragraph{Paper organization.} We begin with preliminaries in
Section~\ref{sec:prelim}. In Section~\ref{sec:building_block},
we discuss the main question of
obtaining a constructive characterization of GS. We prove our main result  in Section~\ref{sec:counterexample},
 that there exists a GS function that is not a positive linear
combination of matroid rank functions.
In Section~\ref{sec:tree},
we present the tree-concordant-sum operation,   show that it preserves
substitutability, use it to show a positive answer to our main question  for $n \leq 4$,  and discuss how it helped in finding the negative instance for the main result. The conclusion is in Section~\ref{sec:conclusion}.

\section{Preliminaries}
\label{sec:prelim}

\paragraph{Valuation functions}
A valuation function is a map $v:2^{[n]} \rightarrow \R$. We restrict
ourselves to functions defined on the hypercube $2^{[n]}$ although the notions
studied here generalize to the integer lattice $\Z^n$.
Given a vector $p \in \R^n$ we define $v_p$ as
the function $$v_p(S) = v(S) - \sum_{i \in S} p_i.$$

\paragraph{Affine transformations and normalized valuations}
We say that a valuation $\tilde{v}$ is an \emph{affine transformation} of $v$ if
there is a vector $p \in \R^n$ and a constant $c \in \R$ such that $\tilde{v} =
c + v_p$. We say that a valuation function $v$ is normalized if
$v(\emptyset) = 0$ and $v(\{i\}) = 0$ for every $i \in [n]$. Every valuation
function can be obtained from a normalized valuation function via an affine
transformation. Unless otherwise specified, we consider matroid rank functions in their normalized form.

\paragraph{Marginals and discrete derivatives}
Given sets $S, T \subseteq
[n]$ we define the marginal contribution of $S$ to $T$ as $v(S \vert T) = v(S \cup T) - v(T)$. We omit parenthesis when
clear from context and often replace $v(\{i,j\}\vert S)$ and $S \cup
\{j\}$ by $v(i,j\vert S)$ and $S \cup j$ respectively.

%It will be also convenient to define the notion of discrete derivative. 
Given a
function $v:2^{[n]} \rightarrow \R$ and $i \in [n]$ we define the derivative
with respect to element $i$ as the function
$\partial_i v:2^{[n]\setminus i} \rightarrow \R$
where $\partial_iv(S) = v(S \cup i) - v(S)$. The first derivative is simply the
marginal $v(i \vert S)$. Applying the operator twice we obtain the second
derivative:
$$\partial_{ij} v(S) = \partial_j [\partial_i v(S)] =
    \partial_i v(S \cup j) - \partial_i v(S) = v(S\cup ij) - v(S \cup i) -
    v(S\cup j) + v(S)$$
%Since we will be mainly concerned with functions where the second derivative is
%non-positive, it will be convenient to define the negative of that quantity so
%that we work with non-negative terms:
%$$\Delta_{ij}^S(v) = -\partial_{ij} v(S).$$
There is a nice economic interpretation of the second derivatives
as a measure of the degree of substitutability of two goods. $\partial_{ij} v(S)$ represents the difference
between the value of the bundle $\{i, j\}$ and sum of values of the two goods $i$ and $j$ separately.
For example, if $\partial_{ij} v(S)= 0$, it means that having good $i$ does not affect the value
for good $j$ conditioned on having a bundle $S$.

\paragraph{Functions as vectors} We often view valuation
functions $v:2^{[n]} \rightarrow \R$ as a vector in $\R^{2^n}$ with coordinates
indexed by $S \subseteq [n]$. This allows
us to view a class of valuation functions as a subset of $\R^{2^n}$. We define
the inner product between two valuations in the usual way:
$$\langle v, w \rangle = \sum_{S \subseteq [n]} v(S) w(S).$$

\subsection{Substitutability}

There are several equivalent ways to define gross substitutes, often also called
GS, discrete concave functions or simply substitutes. The definition that is most convenient to work with for our theorems is the definition via discrete
derivatives due to \citet{ReijnierseGellekomPotters} . The set $\G^n$ of gross substitutes is defined by:
$$\G^n = \{ v:2^{[n]} \rightarrow \R; \text{ } \partial_{ij} v(S) \leq \max
[\partial_{ik}v(S), \partial_{jk}v(S)] \leq 0, \text{ } \forall S \subseteq [n],
\forall i,j,k \notin S  \}$$
We note that the definition does not require monotonicity.

We also state some of the most common ways to define substitutes below. We refer
to \cite{leme2017gross} for a proof that they are equivalent to the definition
above as well as other formulations.

\begin{itemize}
  \item { \bf No price complementarity.} In economics, substitutes was originally
    formulated as the condition that an increase in price for a certain good,
    cannot decrease the demand for other goods. Formally, given a vector $p \in
    R^n$, let $D(v;p) = \argmax_{S \subseteq [n] } v_p(S)$ be the demand
    correspondence. Then $v \in \G^n$ iff for all
    vectors $p \leq p'$ and $S \in D(v;p)$ there is $S' \in D(v;p')$ such that
    $S \cap \{j; p_j = p'_j \} \subseteq S'$.

  \item {\bf Discrete concavity.} In discrete mathematics, substitutes are a
    natural notion of concavity for functions defined in the hypercube. We say
    that a function over the reals $f : \R^n \rightarrow \R$ is
    concave if for every vector $p \in \R^n$, every local minimum of
    $f_p(x) = f(x) - \sum_{i} p_i x_i$ is also a global minimum\footnote{That is
    perhaps not the most common definition of concavity but it is completely
    equivalent to the condition $f(tx +
    (1-t)y) \geq  tf(x) + (1-t) f(y), \forall t \in [0,1]$.}. This definition
    naturally extends to the hypercube:  $v \in \G^n$ iff for every $p \in \R^n$, if
    $S$ is a local minimum of $v_p$, i.e.
    $$v_p(S) \geq v_p(S  \cup i) \quad v_p(S) \geq v_p(S \setminus j) \quad
    v_p(S) \geq v_p(S \cup i \setminus j), \forall i \notin S, j \in S$$
    then $S$ is also a global minimum, i.e. $S \in D(v;p)$.

  \item {\bf Matroidality.} Substitutes can also be defined in terms of greedy
    algorithms: $v \in \G^n$ iff for every $p \in \R^n$,
    the greedy algorithm always computes the maximum of $v_p$, i.e., if we start
    with $S = \emptyset$ and keeps adding the
    element $i \in \argmax_{i} v_p(i \vert S)$ with largest marginal
    contribution while it is positive, we obtain $S \in D(v;p)$.
\end{itemize}

\subsection{Matroids}

Many of those definitions strikingly resemble the
definition of matroids. In fact some of the early
appearances of gross substitutes were attempts to generalize matroids from
collections of subsets to real-valued functions.  \citet{DressTerhalle_WellLayered,DressTerhalle_Rewarding} called
their definition \emph{matroidal maps} and Dress and Wenzel called their
notion \emph{valuated matroids}.

A subset collection over $[n]$
is simply a subset $\M \subseteq 2^{[n]}$. This subset collection is a matroid
if it satisfies one of the following equivalent properties:

\begin{itemize}
  \item {\bf Greedy optimization.} The collection $\M$ is a matroid if for every
    vector $p \in \R^n$, the set $S \in \M$ maximizing $\sum_{i \in S} p_i$ 
    can be obtained by the greedy algorithm
    that starts with the empty set $S = \emptyset$ and keeps adding $i \in
    \argmax_{i; S\cup i \in \M} p_i$ to $S$ while $p_i$ is positive.

  \item {\bf Exchange property.} We say that collection $\M$ is a matroid 
    if $T \subseteq S \in \M$ then $T \in \M$ and for
    every $S, T \in \M$ with $\abs{S} < \abs{T}$, there is $i \in T \setminus
    S$ such that $S\cup i \in \M$.
\end{itemize}

Given any subset system $\M$, we can define its rank function
$r_\M:2^{[n]} \rightarrow \Z_+$ as
$r_\M(S) = \max \{ \abs{T}; T \subseteq S \text{  and } T \in \M \}$. This
allows us to define the set of matroid rank functions as:
$$\MM^n = \{ r_\M; \M \text{ is matroid over } [n] \}$$
When we translate the subset system $\M$ to a rank function $r_\M$, the exchange
property becomes exactly the discrete differential equation in the definition of
$\G^n$. This observation implies that matroid rank functions are exactly the
gross substitutes functions with $\{0,1\}$-marginals:
$$\MM^n = \{v \in \G^n; v(\emptyset) = 0; \partial_i v(S) \in \{0,1\}, \forall S
\subseteq [n], \revision{i \notin S} \} $$
For completeness, we provide a proof of this result in
Theorem~\ref{thm:matroid_submodular} in the appendix.

\subsection{Relation between classes of functions}\label{sec:otherfunctions}

Another class that is important for us is submodular functions:
%, which can be defined as:
$$\SS^n = \{v : 2^{[n]} \rightarrow \R; \partial_{ij}v(S) \leq 0, \forall S
\subseteq [n], i,j \notin S \}$$
%and weighted matroids:
%$$\textstyle \WM^n = \{v : 2^{[n]} \rightarrow \R; \exists 
% w \in \R^n \text{ and } \M \text{ matroid s.t }
%v(S) = \max_{T \in \M, T \subseteq S}
%\sum_{i \in T} w_i, \forall S \subseteq [n]   \} $$
The following relation between the classes hold:
$\MM^n \subseteq  \G^n \subseteq \SS^n$.
The classes $\G^n$ and $\SS^n$ are defined in terms of second order discrete
derivatives, which are invariant under affine transformations, i.e,
if $\tilde{v}$ is obtained from $v$ via an affine
transformation then $\partial_{ij}v(S) = \partial_{ij}\tilde{v}(S)$. In
particular this means that gross substitutes is invariant under affine
transformations. If we want to understand $\G^n$ it is enough to understand the
class $\G^n_0$ of normalized gross substitutes:
$$\G_0^n = \{v \in \G^n; v(\emptyset) = v(i) = 0, \text{ } \forall i \in [n]
\}$$
since we can describe $\G^n = \G_0^n + \E^n$ where $\E^n$ is the class of affine
valuations functions:
$$\textstyle \E^n = \{ v : 2^{[n]} \rightarrow \R; v(S) = c + \sum_{i \in S} p_i; c, p_i \in \R \}$$
It will also be convenient to define the notion of normalized matroid rank
functions:
$$\textstyle \MM^n_0 = \{v : 2^{[n]} \rightarrow \R; \exists \M \text{ matroid s.t. }
v(S) = r_\M(S) - \sum_{i \in S} r_\M(i) \},$$
 which are exactly the normalized gross substitutes functions $\G_0^n$ with $\{-1, 0\}$-marginals.

\section{What are the building blocks of Substitutes
?}\label{sec:building_block}

A major open question in the theory of gross substitutes is how to find a
constructive description of the class. A constructive description has two
parts: a base class of simpler functions and a set of operations that allow us
to build complex functions from simpler ones. There are a number of operations
that are known to preserve substitutability. We mention the two that are most
relevant for this paper here and discuss additional operations in
Section~\ref{sec:conclusion}.

\begin{itemize}
%   \item {\bf Endowment or restriction \cite{HatfieldMilgrom}.}
%     Given a valuation function defined on
%     $n$ items $v:2^{N} \rightarrow \R$ and a subset $S \subseteq N$, we can
%     define a function defined on $N \setminus S$ items as $w:2^{N\setminus S}
%     \rightarrow \R$ such that $w(T) = v(T \vert S)$.
% 
%   \item {\bf Convolution \cite{Murota96} or OR \cite{LehmannLehmannNisan}.}
%     Given two valuation functions $v_1, v_2 : 2^N \rightarrow \R$ we define the
%     convolution $\tilde{v} = v_1 * v_2$ as the function $\tilde{v} : 2^N
%     \rightarrow \R$ such that $\tilde{S} = \max_{T \subseteq S} v_1(T) + v_2(S
%     \setminus T)$.
  \item {\bf Affine transformations. } If $v : 2^{[n]} \rightarrow \R$ satisfies
    gross substitutes and $p \in \R^{[n]}$ is a vector and $u_0 \in \R$ then we can
    build the affine transformation $\tilde{v} : 2^{[n]} \rightarrow \R$ as
    $\tilde{v}(S) = v(S) + \sum_{i \in S} p_i + u_0$.

  \item {\bf Strong Quotient Sum \cite{shioura2012matroid}.} Give two valuations
    $v,w : 2^{[n]} \rightarrow \R$ and $\alpha_1, \alpha_2 \geq 0$
    , we say that $v$ is a strong quotient of $w$ if
    $v(S \vert T) \leq w(S \vert T)$ for all $S, T \subseteq [n]$. Given two
    valuations $v$ and $w$ such that $v$ is a strong quotient of $w$ and $w$ is
    a matroid rank function, we define the strong quotient sum
    $\tilde{v} : 2^{[n]} \rightarrow \R$ as $\tilde{v}(S) = \alpha_1 v(S) + \alpha_2
    w(S)$.
\end{itemize}

Those operations are known to preserve gross substitutability. A major open
question is whether all substitutes can be built from matroid rank functions
using those operations (or perhaps a larger class of simpler operations). We
focus here on affine transformations and positive linear combinations (which is
a strict generalization of the strong quotient sum operation).

\begin{mainquestion}
  Are all gross substitutes positive combinations  of matroid rank functions
  modulo an affine transformation? Formally, given $v \in \G^n$, is there an
  affine transformation $\tilde{v} \in \G^n$, matroid rank functions $r_i \in
  \MM^n$ and positive constants $\alpha_i \in \R_+$ such that:
  $$\textstyle\tilde{v} = \sum_i \alpha_i r_i.$$
  \end{mainquestion}

An equivalent way to ask this question is via the normalized classes: this
allows us to ignore the affine transformations. Given  $v \in \G^n_0$, are
there $r_i \in \MM^n_0$ and $\alpha_i \geq 0$ such that $v = \sum_i \alpha_i
r_i$?

\subsection{Building blocks for submodular functions}

Before we go into our results, we would like to mention a simple
constructive description for submodular functions $\SS^n$ having the set of
matroid rank functions as base. This will serve as a warm up for the study of
substitute valuations. Besides affine transformations, the following
operations preserve submodularity:\footnote{We note that even though positive linear combination and item grouping preserve submodularity, neither of them preserves substitutability in general.}

\begin{itemize}
  \item {\bf Positive linear combination. } If $v_1, v_2:2^{[n]} \rightarrow \R$
    are submodular and $\alpha_1, \alpha_2 \geq 0$ then $\tilde{v} = 
    \alpha_1 v_1 + \alpha_2 v_2$ is also submodular.
  \item {\bf Item grouping.} If $v:2^{[n]} \rightarrow \R$ is submodular and $S_1,
    \hdots, S_k$ is a partition of $[n]$ then the function $w : 2^{[k]}
    \rightarrow \R$ defined as $w(T) = v(\cup_{t \in T} S_t)$ is also
    submodular.
\end{itemize}

It turns out those operations are sufficient to build any submodular function
starting from the set of matroid rank functions. A proof is provided in
Appendix~\ref{appendix:submodular_constr}.

\begin{theorem}\label{thm:submodular_construction}
  Any submodular function can be obtained starting from the set of
  matroid rank functions and applying the operations of
  affine transformations, positive linear combination and item grouping.
\end{theorem}

% !TeX root = main.tex

\section{GS is not in the cone of matroids}
\label{sec:counterexample}

In this section, we exhibit a specific GS function and show
that it cannot be expressed as a positive linear combination of matroid rank
functions. This section is devoted to a (non-computational) proof of that fact.
We defer to Section~\ref{sec:tree} a discussion on how we found such function.
%Thus, despite gross substitutes being a natural generalization of matroid rank
%functions and their multiple similarities, such functions do not suffice to
%provide a constructive description of the class of gross substitutes.  A main
%challenge with this result is to find such a gross substitute function. The
%search space is huge and we find the function using a combination of theory and
%computational techniques. 
%This section is devoted to providing a non-computational  proof that this
%function cannot be written as a positive linear combination of matroid rank
%functions and we defer the discussion of the techniques used to find this
%function to  Section~\ref{sec:tree}.

At a high level, the analysis uses duality and Farkas' lemma. Farkas's
conditions require the existence of a certificate whose inner product with all
matroid rank functions has non-negative sign and the inner product with the
candidate function is strictly negative (Section~\ref{sec:certificate}).
The core of the proof consists in
showing in a non-computational manner that the certificate satisfies the desired
conditions (Section~\ref{sec:conditions}). This is done with a simple lemma
about the local structure of matroid rank functions and a non-trivial partition
of the collection of sets into local groups that can be analyzed individually
with that lemma.

\subsection{The GS function and the certificate}
\label{sec:certificate}

The GS function that we consider for the remaining in this section
is over five elements $[5] = \{1, 2, 3, 4, 5\}$. {This function $v$  is described in its
\emph{normalized form}\footnote{Recall from Section~\ref{sec:prelim} that a
valuation function is normalized if $v(\emptyset) = v(i) = 0$ for all $i \in
[5]$.} in \revision{Figure}~\ref{tab:function}. Since the function is normalized, it is
enough to define it for $\abs{S} \geq 2$.
Checking that the function
satisfies the GS conditions ($v \in \G^5_0$) amounts to checking the
inequalities:
$\partial_{ij}v(S) \leq \max[\partial_{ik}v(S), \partial_{kj}v(S)] \leq 0$
for all $S \subseteq [5]$. There are $40$ such inequalities. It is a tedious
but short verification, which can be found in Appendix \ref{sec:checkcond}.}

Now that we established that $v \in \G^5_0$, we want to prove that there do not
exist (normalized) matroid rank functions\footnote{Recall the
definition of a normalized matroid rank function in
Section~\ref{sec:otherfunctions}.} $r_i \in \MM^5_0$ and $\alpha_i \geq 0$ such
that $v = \sum_{i=1}^n \alpha_i r_i$.

%We consider the
%function $v$ in vector form as a 32-dimensional vector. Let $r_1, \ldots, r_n$
%be the $n$ matroid rank functions over five elements in the same vector form. We
%wish to show that there do not exist $\alpha_1, \ldots, \alpha_n \geq 0$ such
%that $v = \sum_{i=1}^n \alpha_i r_i.$ We begin by recalling Farkas' lemma.

\begin{figure}
\centering
\begin{tabular}{cc|cc|cc}
Sets of size 2 & value & Sets of size 3 & value & Sets of size 4 & value  \\\hline
\{1,2\}        & -1    & \{1,2,3\}      & -2    &  \{1,2,3,4\}              &      -3 \\
\{1,3\}        & -1    & \{1,2,4\}      & -2    &    \{1,2,3,5\}             &-3     \\
\{1,4\}        & 0     & \{1,2,5\}      & -2    &     \{1,2,4,5\}           &      -3 \\
\{1,5\}        & 0     & \{1,3,4\}      & -1    &     \{1,3,4,5\}           &      -2 \\
\{2,3\}        & -1    & \{1,3,5\}      & -1    &      \{2,3,4,5\}            &-2     \\
\{2,4\}        & 0     & \{1,4,5\}      & -1    &     &    \\

\{2,5\}        & 0     & \{2,3,4\}      & -1    &    &    \\
\{3,4\}        & 0     & \{2,3,5\}      & -1    &  Sets of size 5  & value   \\
\cline{5-6}
\{3,5\}        & 0     & \{2,4,5\}      & -1    &    \{1,2,3,4,5\}  &  -4    \\
\{4,5\}        & 0     & \{3,4,5\}      & -1    &   & 
\end{tabular}
\caption{The GS function $v$}
\label{tab:function}
\end{figure}

\begin{lemma}[Farkas' Lemma]
Let $M \in \R^{m\times n}$ and $v \in \R^m$. Then exactly one of the following two statements is true:
\begin{itemize}
\item There exists  $\alpha \in \R^n$ such that $v = M \alpha $ and $\alpha \geq 0$
\item There exists $y \in \R^m$ such that ${M}^{\intercal} y \geq 0$ and $\langle v, y\rangle < 0$.
\end{itemize}
\end{lemma}

We immediately obtain the following corollary which gives two conditions such that, if satisfied, we obtain the desired negative result.

\begin{corollary}
\label{cor:farkas} Let $v$ be a function over five elements. 
If there exists $y \in \R^{32}$ such that  $\langle v, y\rangle < 0$ and $ \langle r, y\rangle \geq 0$ for all normalized matroid rank functions $r$ over five elements,
 then $v$ cannot be expressed as a positive linear combination of normalized matroid rank functions.
\end{corollary}

\subsection{Farkas' conditions}
\label{sec:conditions}

We consider the certificate $y$ given in Figure~\ref{tab:certificate} and show that the two conditions for Corollary~\ref{cor:farkas} hold for that particular certificate $y$ and the gross substitute function $v$. The first and second conditions are shown in Lemma~\ref{lem:farkas1} and Lemma~\ref{lem:farkas2} respectively. The first condition is trivial to verify.

\begin{figure}
\centering
\begin{tabular}{cc|cc|cc}
Sets of size 2 & value & Sets of size 3 & value & Sets of size 4 & value\\\hline
\{1,2\}        & -1    & \{1,2,3\}      & -1    &  \{1,2,3,4\}              &      1  \\
\{1,3\}        & 1    & \{1,2,4\}      & 1    &    \{1,2,3,5\}             &1      \\
\{1,4\}        & -1     & \{1,2,5\}      & 1    &     \{1,2,4,5\}           &      -1  \\
\{1,5\}        & -1     & \{1,3,4\}      & -1    &     \{1,3,4,5\}           &      -1   \\
\{2,3\}        & 1    & \{1,3,5\}      & 1    &      \{2,3,4,5\}            &-1      \\
\{2,4\}        & -1     & \{1,4,5\}      & 1    &     &   \\
\{2,5\}        & -1     & \{2,3,4\}      & -1    &    &    \\
\{3,4\}        & -1     & \{2,3,5\}      & 1    &  Sets of size 5    & value   \\
\cline{5-6}
\{3,5\}        & -1     & \{2,4,5\}      & 1    &  \{1,2,3,4,5\}     & -1   \\
\{4,5\}        & -1     & \{3,4,5\}      & 1    &   &  \end{tabular}
\caption{The certificate $y$}
\label{tab:certificate}
\end{figure}

\begin{lemma} 
\label{lem:farkas1} Let $v$ be the gross substitute function given in Figure~\ref{tab:function} and $y$ be the certificate given in Figure~\ref{tab:certificate}, then
$\langle y, v\rangle < 0$.
\end{lemma}
\proof{Proof.}
This is a simple summation and  we get $\langle y, v\rangle = -1 < 0$.    \\
\endproof

%\begin{table}[]
%\centering
%\label{tab:yv}
%\begin{tabular}{cc|cc|cc|cc}
%Sets of size 2 & value & Sets of size 3 & value & Sets of size 4 & value & Sets of size 5 & value \\\hline
%\{1,2\}        & 1    & \{1,2,3\}      & 2    &  \{1,2,3,4\}              &      -3 &       \{1,2,3,4,5\}         &      4 \\
%\{1,3\}        & -1    & \{1,2,4\}      & -2    &    \{1,2,3,5\}             &-3       &                &       \\
%\{1,4\}        & 0     & \{1,2,5\}      & -2    &     \{1,2,4,5\}           &      3 &                &       \\
%\{1,5\}        & 0     & \{1,3,4\}      & 1    &     \{1,3,4,5\}           &      2 &                &       \\
%\{2,3\}        & -1    & \{1,3,5\}      & -1    &      \{2,3,4,5\}            &2       &                &       \\
%\{2,4\}        & 0     & \{1,4,5\}      & -1    &     &    &                &       \\
%\{2,5\}        & 0     & \{2,3,4\}      & 1    &    &     &                &       \\
%\{3,4\}        & 0     & \{2,3,5\}      & -1    &     &     &                &       \\
%\{3,5\}        & 0     & \{2,4,5\}      & -1    &     &   &                &       \\
%\{4,5\}        & 0     & \{3,4,5\}      & -1    &   &     &   &     
%\end{tabular}
%\caption{$y^\intercal \v = -1$}
%\end{table}

The interesting condition to show is  $\langle y, r\rangle \geq 0$ for all matroid rank functions $r$. We first give a simple lemma about the local structure of matroid rank functions which will motivate the approach for the analysis of this second condition.

\begin{lemma}
\label{lem:matroid}
Let $r$ be a matroid rank function. For any set $S$ and elements $a_1, a_2 \not \in S$, 
\begin{itemize}
\item if  $r(S \cup a_1) - r(S) = - 1$ and $ r(S \cup a_2) - r(S) = -1$, then
  $r(S \cup \revisiontwo{\{a_1, a_2\}}) - r(S) = - 2$;
\item if  $r(S \cup a_1) - r(S) = -1 $ and $ r(S \cup a_2) - r(S) = 0$, then
  $r(S \cup \revisiontwo{\{a_1 , a_2\}}) - r(S) = - 1$.
\end{itemize}
\end{lemma}
\proof{Proof.}
We first decompose the quantity of interest in two terms,
\begin{align*}
r(S \cup a_1 , a_2) - r(S)  = \left(r(S \cup a_1 , a_2) - r(S \cup a_2)\right) +
  \left(r(S \cup a_2) - r(S)\right).
\end{align*}
Since $r$ is a matroid rank function, marginal contributions are either $-1$ or
  $0$ and $r(S \cup a_1 , a_2) - r(S \cup a_2)  \in \{-1, 0\}$. Next, by
  submodularity and the assumption for both cases, we get $r(S \cup a_1 , a_2) - r(S \cup a_2) \leq r(S \cup a_1) - r(S)
  = -1$. Thus,  $r(S \cup a_1 , a_2) - r(S \cup
  a_2) = -1$ and 
$$r(S \cup a_1 , a_2) - r(S)  = -1 + r(S \cup a_2) - r(S),$$
which concludes the proof.  
\endproof

The main idea to show that $\langle y, r \rangle \geq 0$ for all matroid rank functions $r$ is to first partition the sets into six local groups $G_1, \ldots, G_6$, described in Figure~\ref{tab:partition}, and then use Lemma~\ref{lem:matroid} to argue about the value $\langle y_G, r_G\rangle$ for each local group $G$, where $v_G$ denotes the subvector of a vector $v$ of length $|G|$ induced by the indices corresponding to sets in group $G$. %To argue about each group at a local level, we give the following lemma about the local structure of matroid rank function. This is a simple yet very useful lemma which will be used multiple times to show the desired condition.

\begin{figure}
\centering
\begin{tabular}{cc|cc|cc|cc}
Group 1 & value & Group 2   &  value  & Group 4     &  value  & Group 5   & value    \\\hline
\{3,4\} & -1    & \{1,3\}   & 1  & \{1,5\}     & -1 & \{1,2\}    & -1   \\
\{4,5\} & -1    & \{1,4\}   & -1 & \{2,5\}     & -1 & \{1,2,4\}   & 1   \\
        &       & \{1,3,4\} & -1 & \{3,5\}     & -1 & \{1,2,5\}   & 1   \\
        &       &           &    & \{1,4,5\}   & 1  & \{1,2,4,5\}  & -1  \\
        &       & Group 3   & value   & \{2,4,5\}   & 1  &       &        \\
        &       & \{2,3\}   & 1  & \{3,4,5\}   & 1  & Group 6   &  value   \\
        &       & \{2,4\}   & -1 & \{2,3,5\}   & 1  & \{1,2,3\}  & -1    \\
        &       & \{2,3,4\} & -1 & \{2,3,4,5\} & -1 & \{1,2,3,4\}  & 1  \\
        &       &           &    & \{1,3,5\}   & 1  & \{1,2,3,5\}   & 1 \\
        &       &           &    & \{1,3,4,5\} & -1 & \{1,2,3,4,5\}& -1 
\end{tabular}
\caption{The partition of the sets into groups, \revision{with the corresponding values of $y$.}}
\label{tab:partition}
\end{figure}

%We are now ready to show the second condition for Corollary~\ref{cor:farkas}.
\begin{lemma}
\label{lem:farkas2}
Let $y$ be the certificate. Then for all matroid rank functions $r$,  $\langle y, r\rangle \geq 0$.
\end{lemma}
\proof{Proof.}
 We first show that for all $G_i$ such that $i \neq 4$,  $\langle y_{G_i},
 r_{G_i}\rangle \geq 0$ for all matroid rank functions $r$. Then, we show that
 $\langle y_{G_4}, r_{G_4}\rangle \geq -1$ and that if $\langle y_{G_4},
 r_{G_4}\rangle = -1$, then for at least one other group $G$, $\langle y_{G},
 r_{G} \rangle \geq 1$. Since $\langle y, r \rangle = \sum_{i =1}^6 \langle
 y_{G_i}, r_{G_i}\rangle$, we then obtain  $\langle y, r \rangle \geq 0$ (recall
 that the empty set and singletons are normalized to have value $0$
 \revision{and that function values are nonpositive for all subsets}).

We first show that for all $G_i$ such that $i \neq 4$,  $\langle y_{G_i}, r_{G_i}\rangle \geq 0$ for all matroid rank functions $r$. This is trivial for group $G_1$ since $r(S) \leq 0$ for all sets $S$. For group $G_2$, recall that  we have 
$$y(1, 3) = 1 \ \ \ \ y(1, 4) = -1 \ \ \ \  y(1, 3, 4) = -1.$$ 
Since $r(1) = 0$ and the marginal contributions of matroid rank functions
\revision{are either $0$ or $-1$,} $r(1, 3), r(1, 4) \in \{-1,0\}$ and we consider the following three cases:
\begin{itemize}
\item If $r(1, 3)= -1$ and $r(1, 4) = -1$, then by Lemma~\ref{lem:matroid}, $r(1,3,4) = -2$. We get $\langle y_{G_2}, r_{G_2}\rangle = 2$. 
\item If $r(1, 3) = -1$ and $ r(1, 4) = 0$, then by Lemma~\ref{lem:matroid}, $r(1,3,4) = -1$ and we  get $\langle y_{G_2}, r_{G_2}\rangle = 0$. 
\item If  $r(1, 3) = 0$, then  $\langle y_{G_2}, r_{G_2}\rangle \geq 0$. 
\end{itemize}

Group $G_3$ follows similarly as for group $G_2$. For group $G_5$, recall that we have 
$$y(1, 2) = -1 \ \ \ \  y(1, 2,4) = 1 \ \ \ \ y(1, 2,5)=1 \ \ \ \ y(1, 2,4,5) = -1.$$ 
%There are four cases dependent on $r(1,2, 4) - r(1, 2), r(1,2, 5) - r(1, 2) \in \{-1,0\}$: 
%\begin{itemize}
%\item If $r(1,2, 4) - r(1, 2) = 0$ and $r(1,2, 5) - r(1, 2) = 0$, then   $y_{G_5}^{\intercal} r_{G_5} \geq 0$ since $r(1, 2,4,5) \leq  r(1, 2), r(1,2, 4), r(1,2, 5) $.
%\item If $r(1,2, 4) - r(1, 2) = 0$ and  $r(1,2, 5) - r(1, 2) = -1$, then by Lemma~\ref{lem:matroid},  $r(1, 2,4,5) -  r(1,2, 4) = -1$ and we get $y_{G_5}^{\intercal} r_{G_5} = 0$.
%\item If $r(1,2, 4) - r(1, 2) = -1$ and  $r(1,2, 5) - r(1, 2) = 0$, then $y_{G_5}^{\intercal} r_{G_5} = 0$ similarly as for the previous case. 
%\item If $r(1,2, 4) - r(1, 2) = -1$ and  $r(1,2, 5) - r(1, 2) = -1$, then by Lemma~\ref{lem:matroid}, $r(1, 2,4,5) -  r(1,2, 4) = -2$ and we get $y_{G_5}^{\intercal} r_{G_5} = 0$.
%\end{itemize}
Since $r$ is submodular, $r(S) + r(T) \geq r(S \cap T) + r(S \cup T)$ for any sets $S,T$. Thus, $-r(1, 2) + r(1, 2,4) + r(1, 2,5) - r(1, 2,4,5) \geq 0$ and this implies that $\langle y_{G_5}, r_{G_5}\rangle \geq 0$.
Group $G_6$ follows similarly as for group $G_5$.

It remains to show  that  $\langle y_{G_4}, r_{G_4}\rangle \geq -1$ and that if $\langle y_{G_4}, r_{G_4}\rangle = -1$, then for at least one other group $G$, $\langle y_{G}, r_{G}\rangle \geq 1$. First, consider the following three possible partitions of $G_4$.
\begin{align*}
\left( \{3, 5\}, \{3,4,5\}, \{1, 3,5\}, \{1,3,4,5\}\right), \left(\{ 2, 5\}, \{2, 4,5\}, \{2, 3,5\}, \{2, 3, 4,5\}\right),  \left(\{1, 5\}, \{1,4,5\}\right) \\
\left( \{3, 5\}, \{3,4,5\}, \{2, 3,5\}, \{2,3,4,5\}\right),
\left(\{ 1, 5\}, \{1, 4,5\}, \{1, 3,5\}, \{1, 3, 4,5\}\right),  \left(\{2, 5\}, \{2,4,5\}\right) \\
 \left(\{2, 5\}, \{2,4,5\}, \{2, 3,5\}, \{2,3,4,5\}\right),
\left(\{ 1, 5\}, \{1, 4,5\}, \{1, 3,5\}, \{1, 3, 4,5\}\right), \left(\{3, 5\}, \{3,4,5\}\right) 
\end{align*}
For the first two parts $G_4'$ and $G_4''$ of each possible partition, similarly as for $G_5$, we obtain that $\langle y_{G_4'}, r_{G_4'}\rangle \geq 0$ and $\langle y_{G_4''}, r_{G_4''}\rangle \geq 0$. For the last part $G'''_5$, it is easy to see that $\langle y_{G_4'''}, r_{G_4'''}\rangle \geq -1$. Thus, we obtain that  $\langle y_{G_4}, r_{G_4}\rangle \geq -1$.

Next, consider 
$$r(1,4,5) - r(1, 5) \ \ \ \ r(2,4,5) - r(2, 5) \ \ \ \ r(3,4,5) - r(3, 5).$$

If one of these three difference is $0$, then by considering the corresponding above partition where the two terms in the difference form the last part $G_4'''$ of the partition, we get that $\langle y_{G_4}, r_{G_4}\rangle \geq 0$.

Next, consider the case where these three differences are all equal to $-1$ and
$\langle y_{G_4}, r_{G_4}\rangle = -1$. If $r(4,5) = - 1$, then $y(4,5) r(4,5) =
1$ compensates for $\langle y_{G_4}, r_{G_4}\rangle = -1$ and $\langle y, r
\rangle \geq 0$. Otherwise, $r(4,5) = 0$. This implies  \revision{ that  $r(1,4,5) = r(2,4,5) = r(3,4,5) = -1$ and $r(1,5) = r(2,5) = r(3,5) = 0$} since the above differences are all $-1$. It must also be the case that $r(2,3,4,5) = -2$ by submodularity since $a_2$ has marginal contribution $-1$ to $\{a_4, a_5\}$ and since $ r(3,4,5) = -1$. Similarly, $r(1,3,4,5) = -2$.
 
 Next, we  focus on $G_2$ and recall that we have 
 $$y(1, 3) =  1 \ \ \ \ y(1, 4) = -1 \ \ \ \ y(1, 3, 4) = -1.$$ 
 Note that $r(1,3,4) \leq -1$ since $r(1,3,4,5) = -2$. Since $r(1,3) \geq r(1,3,4)$, it must be the case that $r(1,4) = 0$, $r(1,3) \revisiontwo{ = } r(1,3,4) = -1$ for $\langle y_{G_2}, r_{G_2}\rangle = 0$. If that is not the case, then $\langle y_{G_2}, r_{G_2}\rangle \geq 1$ and that compensates for $G_4$ and we get $\langle y, r \rangle \geq 0$.
 
 Thus, in the remaining case $r(1,3) = -1$.  Similarly for $G_3$ with $r(2,3,4,5) = -2$, the remaining  case is if $r(2,4) = 0$ and  $r(2,3) = -1$. Next,  with $r(2,3) = r(1,3) = -1$, then, by Lemma~\ref{lem:matroid}, $r(1,2,3) = -2$, which in turn implies that $r(1,2) = -1$.
 
 Next, we consider $G_5$. Recall that 
$$y(1, 2) = -1 \ \ \ \  y(1, 2,4) = 1 \ \ \ \ y(1, 2,5)=1 \ \ \ \ y(1, 2,4,5) = -1.$$
Since  $r(4,5) =  0$, $r(1,4,5) = -1$ and $r(2,4,5) = -1$, by Lemma~\ref{lem:matroid},  $r(1,2,4,5) = -2$. If $r(1,2,5) \revisiontwo{ = }  r(1,2,4) = - 1$, then  with $r(1,2) = -1$, $\langle y_{G_5}, r_{G_5}\rangle \geq 1$ and we are done. Otherwise, $r(1,2,5)$ or $r(1,2,4)$ is equal to  $- 2$. But this is impossible since we are in a case where $r(1,5) = 0$ and $r(1,4) = 0$. Thus, if  $\langle y_{G_4}, r_{G_4}\rangle = -1$, then  for at least one other group $G$, $\langle y_{G}, r_{G}\rangle \geq 1$ and we get $\langle y, r \rangle \geq 0$.  
\endproof

Combining Corollary~\ref{cor:farkas}, Lemma~\ref{lem:farkas1} and  Lemma~\ref{lem:farkas2}, we obtain the main result.

\begin{theorem}
\label{thm:main}
The gross substitute function $v$ cannot be expressed as a positive linear
  combination of normalized matroid rank functions. {In particular, no affine
  transformation of $v$ can be expressed as a positive linear combination of
  matroid rank functions.}
\end{theorem}

In particular, this implies that the strong quotient sum and \revision{ tree-concordant-sum} operations are not sufficient to construct all gross substitutes from matroid rank functions and that MRS valuations do not exhaust gross substitutes. We note that even though $v$ is non-monotone, the main result holds for monotone gross substitute functions since there exists some affine transformation of $v$ that is monotone.
We also extend the negative result from matroid rank functions to weighted matroid rank functions. This follows from the fact that weighted matroid rank functions can be expressed as a positive linear combination of unweighted matroid rank functions, which was proven in \cite{shioura2012matroid} and we give a proof in Appendix~\ref{sec:appcounterexample} for completeness as Lemma~\ref{lem:weighted}.

\begin{corollary} {No affine transformation of function $v$ can} be expressed as a positive linear combination of weighted matroid rank functions.
\end{corollary}

\section{The Tree-Concordant-Sum Operation}
\label{sec:tree}

In this section, we define the notion of a tree representation which
abstracts the combinatorial structure of \emph{tree-form Hessians} of
\citet{hira2004m}. We first  show that this
representation has the following nice property: the condition that two functions
have such a tree representation that is \emph{compatible} is necessary and sufficient
for the summation operation to preserve substitutability. We call this new
operation preserving substitutability tree-concordant-sum and show that it also
provides a polyhedral characterization of gross substitutes.

We then use this representation to give a positive answer to our main question
for $n \leq 4$: a GS over at most $4$ elements can be
written as a positive linear combination of matroid rank functions. This implies
that at least $5$ elements are necessary to obtain the negative answer from the
previous section.

Finally, we discuss how the polyhedral understanding of gross substitutes based
on this tree representation combined with computational techniques led to the
discovery of the counterexample for $n= 5$.

\subsection{Substitution Trees}

The tree representation of GS is best explained using the
discrete derivative property introduced in Section~\ref{sec:prelim}. Since it
will be convenient to work with non-negative numbers, we introduce the notation:
$$\Delta_{ij}^S(v) = -\partial_{ij} v(S)$$
omitting $v$ when clear from context. In this notation,
we can write the GS condition as:
 $$\Delta^S_{ij} \geq \min(\Delta^S_{ik}, \Delta^S_{jk}) \geq 0$$
 If we permute the identities of
$i,j,k$ such that the symbols are sorted, we have the following triangle
property: $\Delta^{S}_{ij} = \Delta^S_{ik} \leq \Delta^S_{jk}$.
\citet{hira2004m} and  \citet{BingLehmannMilgrom}
note that this resembles the definition of \emph{ultra-metrics}, which admit
tree-like representations. This enables similar tree-like representations for
GS. Below we describe the notion of \citet{hira2004m}
following the presentation in~\revisiontwo{\citet{leme2017gross}}.

\begin{theorem}[\citet{hira2004m}]\label{thm:hirai}
  A function $v$ satisfies the GS condition iff for every subset $S$ there is a
  tree $T(v)^S$ having the elements of $[n] \setminus S$ in the leaves and
  non-negative real number labels in the internal nodes such that:
  \begin{itemize}
    \item The label of each internal node is larger than or equal to the label of its
      parent.
    \item For every $i,j \notin S$, $\Delta^S_{ij}$ corresponds to the label of
      the lowest common ancestor $(\lca)$ of the leaves corresponding to $i$ and $j$.
  \end{itemize}
\end{theorem}

We observe that the representation of Murota and Hirai has two components: a
purely combinatorial structure, which is the collection of trees and a numerical
component, which are the values of the labels. If we abstract the numerical
component, we obtain what we call a tree structure:

\begin{definition}[Tree structure] A tree structure corresponds to a collection
  of trees $\{T^S\}_S$ indexed by subsets $S \subseteq [n]$ such that the leaves
  of tree $T^S$ correspond to the elements of $[n] \setminus S$.
\end{definition}

\begin{figure}
\centering
\begin{subfigure}{.25\textwidth}
  $\left[ \begin{aligned} 
    & * && 2 && 1 && 1 && 1 \\
    & 2 && * && 1 && 1 && 1 \\
    & 1 && 1 && * && 3 && 3 \\
    & 1 && 1 && 3 && * && 3 \\
    & 1 && 1 && 3 && 3 && * \\
  \end{aligned} \right] \quad$
\end{subfigure}
\begin{subfigure}{.35\textwidth}
  \begin{tikzpicture}[scale=.6]
  \tikzstyle{level 1}=[sibling distance=30mm]
  \tikzstyle{level 2}=[sibling distance=15mm]
	\node[circle, draw=black]{$1$}
    child {node[circle, draw=black] {$2$}
      child {node {$a$}}  
      child {node {$b$}}  
    }
    child {node[circle, draw=black] {$3$}
       child {node {$c$}}  
       child {node {$d$}}
       child {node {$e$}}
    };
\end{tikzpicture}
\end{subfigure}%
\begin{subfigure}{.35\textwidth}
  \begin{tikzpicture}[scale=.6]
  \tikzstyle{level 1}=[sibling distance=30mm]
  \tikzstyle{level 2}=[sibling distance=15mm]
  \tikzstyle{level 1}=[sibling distance=30mm]
  \tikzstyle{level 2}=[sibling distance=15mm]
	\node[circle, draw=black]{$1$}
    child {node[circle, draw=black] {$2$}
      child {node {$a$}}  
      child {node {$b$}}  
    }
    child {node[circle, draw=black] {$3$}
      child {node[circle, draw=black] {$3$}
       child {node {$c$}}
       child {node {$d$}}
      }
      child {node {$e$}}
    };
\end{tikzpicture}
\end{subfigure}
  \caption{Two valid tree representations for the same matrix $[\Delta^S_{ij}]_{ij}$.}
\label{fig:treepartition}
\end{figure}
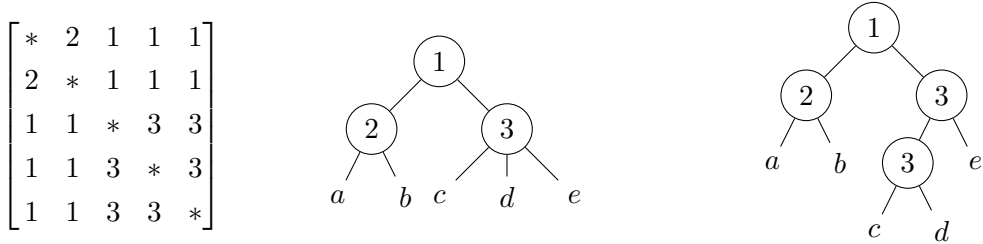

We say that a valuation function $v$ admits a tree structure $\{T^S\}_S$ if
we can represent $v$ in the sense of Theorem~\ref{thm:hirai} using those trees.
The tree structure might not be unique.
For example, if for a certain $v$ and $S$,
$\Delta_{ij}^S$ is given by the matrix in the left of Figure
\ref{fig:treepartition}, then the two trees in the figure are valid
structures for $\Delta_{ij}^S$. There is therefore some flexibility in the
choice of the tree structure, which allow us to define the notion of
tree-concordant:

\begin{definition}[Tree-concordance]
  We say that two GS valuations are \emph{tree concordant} if they admit the
  same tree structure $\{T^S\}_S$.
\end{definition}

A clean way to check when two functions are tree concordant is via the concept
of minimal representation. We say that a tree representation for
valuation $v$ is \emph{minimal} if no node has the same label as its parent. The tree
on the left in Figure~\ref{fig:treepartition} is minimal, for example while the
one on the right is not. By the definition of minimal, it is clear that each GS
valuation has an unique minimal representation. It can be obtained by starting
from any representation and collapsing tree edges connecting internal nodes with
the same label.

%%%%%%%%%%%%%%%%

To check when there is one tree structure that two functions simultaneously
admit, it is enough to look at the minimal representations. To see that, it is
useful to view a tree as a laminar family. Given a tree $T^S$ with elements $[n]
\setminus S$ in the leaves, we can represent it by a family of subsets $L^S$
constructed as follows: a set $X \subseteq [n] \setminus S$ is in $L^S$ \revisiontwo{if and only if} there
is an internal node \revisiontwo{v} in $T^S$ such that $X$ is the set of leaves below $v$. Such subset
collection is what is called a laminar family:

\begin{definition}[Laminar family] \revision{A collection $L$ of subsets } is called a
  laminar family if for every $X, Y \in L$ either: (i) $X \cap Y = \emptyset$;
  or (ii) $X \subseteq Y$ or (iii) $Y \subseteq X$.
\end{definition}

Now, we can check if two functions $u$ and $v$ are tree-concordant as follows:

\begin{lemma}\label{lemma:laminar}
  If $u$ and $v$ are GS functions and $\{T^S(u)\}$ and $\{T^S(v)\}$
  are its minimal tree structures and $\{L^S(u)\}$ and $\{L^S(v)\}$ are its
  corresponding laminar family representations, then $u$ and $v$ are tree
  concordant iff for every $S$, $L^S(u) \cup L^S(v)$ is also a laminar family.
\end{lemma}

The proof is straightforward from definitions and the fact that there is natural
one-to-one mapping between trees and laminar families. What is
interesting  about tree concordance is that it provides necessary and
sufficient conditions for the sum of two GS functions to be GS:

\begin{theorem}
  Let $u$ and $v$ be two gross substitute functions. The function
  $u + v$ is a gross substitute function if and only if $u$ and $v$ are
  tree-concordant.
\end{theorem}
\proof{Proof.}
  We first show that if $u$ and $v$ are tree-concordant, then $u+v$ is a
  gross-substitute. Consider the tree representation $T$ that has the structure
  of $u$ and $v$, which is identical since they are tree-concordant, but with
  numerical values at each internal node that is the sum of the values at that
  node for $u$ and $v$. Since second order derivatives are linear:
  $$ \Delta_{ij}^S(u+v) = \Delta_{ij}^S(u) + \Delta_{ij}^S(v)  $$
  and it clearly admits the same tree structure $T$, so by
  Theorem~\ref{thm:hirai}, $u+v$ is in GS.

  Next, we show that if $u$ and $v$ are not tree-concordant, then $u+v$ is not a
  gross substitute function. Assume $u$ and $v$ are not tree-concordant
  then by Lemma~\ref{lemma:laminar} there is a set $S$ such that for the laminar
  representations $L^S(u)$ and $L^S(v)$ there are sets with non-trivial
  intersection, i.e., there are $X \in L^S(u)$ and $Y \in L^S(v)$ such that
  we can find $i \in X \setminus Y$, $j \in X \cap Y$ and $k \in Y \setminus X$.
  Therefore:
  $$\Delta^S_{ij}(u) > \Delta^S_{ik}(u) = \Delta^S_{kj}(u)$$
  $$\Delta^S_{ij}(v) = \Delta^S_{ik}(v) < \Delta^S_{kj}(v)$$
  therefore, it must be the case that:
  $$ \Delta^S_{ik}(u+v) < \min[ \Delta^S_{ij}(u+v), \Delta^S_{kj}(u+v) ]$$
  which is a violation of gross substitutability.
\endproof

Since it is a necessary and sufficient condition, this includes the
strong-quotient-sum property in \cite{shioura2012matroid} as a special case:

\begin{corollary}
  If a GS function $u$ and a matroid rank function $v$ satisfy the 
  strong-quotient-sum property, then they are tree-concordant.
\end{corollary}

\subsection{Polyhedral description of $\G^n$}

The concept of tree structure and tree concordance provide a good tool for
describing the geometry of $\G^n$. Viewing valuations as vectors in $\R^{2^n}$,
we can view $\G^n$ as a subset of that space. Lehmann, Lehmann and Nisan
\cite{LehmannLehmannNisan} observe that the set is a non-convex and has zero
measure.

The concepts developed earlier in this section allow us to decompose
the space in finitely many convex cones. Since a tree structure is a finite
combinatorial object, there are finitely many such structures. Fix a tree
structure $\{T^S\}_S$ and consider the subset of $\G^n$ with functions that
admit $\{T^S\}_S$. All functions in this subset are tree concordant, so the set
is closed under positive linear combination, forming a \emph{convex cone}.

What we would like to do next is to understand how those cones look like. For
that we will fix a tree structure and assign labels to internal nodes. The
labels must be non-negative and the label of a node cannot be smaller than the
label of its parent, but those conditions alone are not sufficient for the
existence of a valuation function producing those $\Delta_{ij}^S$. An extra
condition that is required is what we call \emph{integrability}\footnote{We
thank Kazuo Murota for his suggestion on a earlier version of this manuscript to
phrase this analysis in terms of integrability condition.}\revisiontwo{(we simplify the notation for $S \cup \{i\}$ with $S+i$ for the remainder of the paper)}:

\begin{equation}\label{eq:int}\tag{Int}
  \Delta^{\revision{S + k}}_{ij} - \Delta^S_{ij} = \Delta^{\revision{S + j}}_{ik} - \Delta^S_{ik}
  = \Delta^{\revision{S + i}}_{jk} - \Delta^S_{jk}.
\end{equation}

\begin{lemma}[Integrability]
  Given values $\Delta^S_{ij}$ \revision{for all $i, j, S$ such that $i, j \not \in S$,} there is a valuation $v$ such that\
  $\partial_{ij}v(S) = - \Delta^S_{ij}$ \revision{if and only if} the integrability conditions
  (\ref{eq:int}) hold.
\end{lemma}

In Appendix~\ref{appendix:integrability} we provide a discussion on
integrability conditions for discrete functions. The lemma above follows
directly from Lemma~\ref{lemma:int:2} in the appendix.

Since the symbols $\Delta_{ij}^S$ have a special tree-form for gross
substitutes, it is convenient to re-write the integrability conditions in the
following form. 

\begin{lemma}\label{lem:welldefined} An assignment of labels to the internal
  nodes of a tree structure corresponds to a representation of GS function
  \revision{if and only if}
  the following condition holds for every $S \subseteq [n]$ and $i,j,k \notin
  S$:
  $$\begin{aligned}
    & \text{ if }  & &\Delta_{ik}^S = \Delta_{jk}^S = \Delta_{ij}^S - \alpha,
    \revision{\text{ for some } \alpha \geq 0 }\\
    &  \text{ then } & &
  \Delta_{ik}^{S+j} = \Delta_{jk}^{S+i} = \Delta_{ij}^{S+k} - \alpha
  \end{aligned}$$
\end{lemma}

The following corollary (which is a rephrasing of Corollary~\ref{cor:int:3}
in the appendix) shows how to explicitly reconstruct a function from second order 
derivatives. \revision{We denote by $S_{<i}$ the set of elements smaller than $i$ according to their label, i.e., $S_{<i} = \{j : j < i\}$.}

\begin{corollary}\label{cor:integration}
  Given $\Delta_{ij}^S$ satisfying the integrability conditions (\ref{eq:int}),
  the unique normalized function such that $\partial_{ij} v(S) =
  -\Delta_{ij}^S$ is given by:
  $$v(S) = -\sum_{i,j \in S; i < j} \Delta_{ij}^{S_{<i}}$$
  In particular, all the functions such that $\partial_{ij} v(S) =
  -\Delta_{ij}^S$ are
  affine transformations of the function defined above.
\end{corollary}

\subsection{Representations of $\G$}

Using the tree representation of gross substitutes, we  provide a constructive
characterization of gross substitutes $\G^n$ for $n \leq 4$ from matroids.
Given a  matroid $\M$ we denote its rank by $r[\M]$.
We  denote the uniform matroid of rank $i$ over $j$ elements by $U^i_j$. 

Given two functions $v$ and $\tilde{v}$ we say that they are equivalent (up to
affine transformations) if $v - \tilde{v} \in \E^n$. Given a normalized
valuation $v \in \G^n_0$, we will often associate it with the equivalent
function $\tilde{v}(S) = v(S) + \abs{S}$. Often, a normalized matroid rank
function $v$ is easier to recognize in its $\tilde{v}$ form.

Moreover, we write $X \simeq Y $ to denote a linear isomorphism between two
sets, i.e., if there is a linear bijection $L$ such that $L(X) = Y$.

\subsubsection{Description of $\G^2$}

For $n = 2$, the only constraint is $\Delta^\emptyset_{1,2} \geq 0$ so:
$$\revision{\G^2_0 = \{(v(\emptyset) = 0, v(1) = 0, v(2) = 0, v(12)) ; v(12) \leq 0 \} \simeq \R_+.}$$
Thus, $v=(0,0,0,-1)$ is a representative of the class $\G_0^2$ and $\tilde{v}= (0,1,1,1)$, which is the rank function $r[U_2^1]$ of the uniform
matroid of rank $1$ over $2$ elements,   is a representative \revision{of the class $\G^2$}.  Since $\E^2 = (1,1,1,1) \cdot \R +  (0,1,0,1) \cdot 
\R + (0,0,1,1) \cdot \R \simeq
\R^3$, we have:
$$\G^2 = \E^2 + r[U_2^1] \cdot \R_+ \simeq \R^3 \times \R_+.$$
 The set $\G^2$ is not very interesting since the set of gross substitutes on $2$
variables is the same as the set of submodular functions on $2$ variables, which
is known to be a convex set.

%\comment{
%In fact, let the $\S^n$ be the set of submodular functions on $n$ variables and
%let $\S^n_0 := \{v \in \S^n; v(\emptyset) = 0, v(i) = 0, \forall i \in N\}$.
%Then by the same argument (that submodular functions depend only on the second
%order derivatives), we have $\S^n = \S^n_0 + \E^n$. Now, the formula
%$v(S) = -(n-1) \cdot v(\emptyset) + \sum_{i \in S} v(i) - \sum_{i<j}
%\Delta_{ij}^{S_{<i}}$ establishes a linear bijection between $\S^n_0$ and
%$\R^{2^n-(n+1)}_+$, via the mapping from $(\Delta_{ij}^{S_{<i}})_{i<j, S
%\subseteq[i-1]}$ to $(v(S))_{S \subseteq N} \in \S^n_0$. So:
%$$\S^n \simeq \R^{n+1} \times \R^{2^n-(n+1)}_+$$
%}

\subsubsection{Description of $\G^3$}

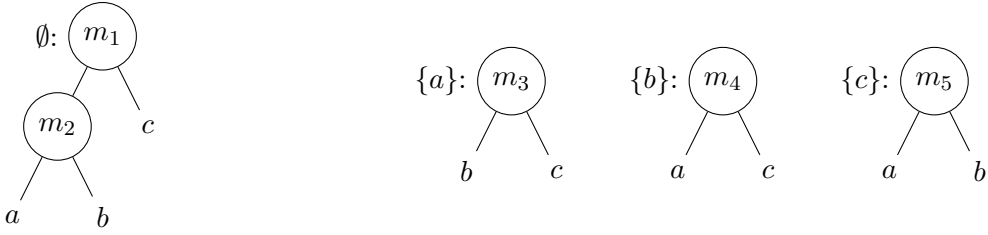
\begin{figure}
\centering
\begin{subfigure}{.3\textwidth}
  \centering
  \begin{tikzpicture}[scale=.8]
  \node[circle, draw=black, label=left:$\emptyset$:] {$m_1$}
    child {node[circle, draw=black] {$m_2$}
      child {node {$a$}}
      child {node {$b$}}}
		child {node {$c$}};
\end{tikzpicture}
\end{subfigure}%
\begin{subfigure}{.7\textwidth}
  \centering
  \begin{tikzpicture}[scale=.8]
  \node[circle, draw=black, label=left:$\{a\}$:] {$m_3$}
      child {node {$b$}}
      child {node {$c$}};
  \node[xshift=80pt, circle, draw=black, label=left:$\{b\}$:]{$m_4$}
      child {node {$a$}}
      child {node {$c$}};
  \node[xshift=160pt, circle, draw=black, label=left:$\{c\}$:]{$m_5$}
      child {node {$a$}}
      child {node {$b$}};
\end{tikzpicture}
\end{subfigure}
\caption{The tree structure for functions in $\G^3$.}
\label{fig:G32}
\end{figure}

The set of gross substitutes on $3$ items is more interesting since it is not
convex. We name the items $\{a,b,c\}$.
For every $v \in \G^3_0$ and up to the renaming of the elements, there is only one possibility for the
substitution tree associated to the empty set, which is depicted in Figure~\ref{fig:G32}. For each singleton set, there is also only one possible
substitution tree. Applying Lemma~\ref{lem:welldefined} that is required to
obtain a well-defined function, with $S = \emptyset$, we obtain the following
additional necessary and sufficient constraint \revision{for the labels in Figure~\ref{fig:G32}}:
$$ m_3 = m_4 = m_5 - (m_2 - m_1).$$

%\caption{.}

%%%%%%%%%%%%%%%%

We  write $m_1 = x$, $m_2 - m_1 = y$ (recall $m_2 \geq m_1$) and $m_3 = z$, so we can parametrize the space
of feasible $m = (m_1, m_2, m_3, m_4,m_5)$ by:
$$(x,y,z) \in \R^3_+ \mapsto m = (x, x+y, z, z, y+z)$$
In other words, the space of feasible values of $m$ are a cone generated by the
vectors \\
$(1,1,0,0,0), (0,0,1,1,1), (0,1,0,0,1)$. It is particularly interesting
to see which valuations they correspond to. 

\paragraph{Vector $(1,1,0,0,0)$} By solving the equation to obtain $v$ from
second derivatives $\Delta$ (Corollary \ref{cor:integration}), we obtain the valuation $v \in \G^3_0$ such that
$v(ab) = v(bc) = v(ac) = -1$ and
$v(abc) = -2$.   Thus, $\tilde{v}(S) = v(S) + \abs{S} = 1$ for all $S \neq \emptyset$, which is the rank
function $r[U_3^1]$.

\paragraph{Vector $(0,0,1,1,1)$} Solving the equations for $v$ we obtain $v(ab)
= v(bc) = v(ac) = 0$ and $v(abc) = -1$, so $\tilde{v}(S) = \min\{
  \abs{S}, 2 \}$ which is  $r[U_3^2]$.

\paragraph{Vector $(0,1,0,0,1)$} More interestingly, by solving for $v$ we
obtain: $v(ac) = v(bc) = 0$ and $v(ab) = v(abc) = -1$, so $\tilde{v}$ is such that the singletons have value $1$, $\tilde{v}(ab) =
1$ and $\tilde{v}(bc) = \tilde{v}(ac) = \tilde{v}(abc) = 2$. This is the
rank function of the graphical matroid associated with the following graph:

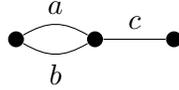
\begin{center}
  \begin{tikzpicture}[scale=.7]
  \tikzstyle{vertex}=[circle,fill=black,minimum size=6pt,inner sep=0pt];
  \node[vertex] (a) at (0,0) {};
  \node[vertex] (b) at (1.5,0)  {};
  \node[vertex] (c) at (3,0) {};
  \draw (a) .. controls +(-30:20pt) and +(-150:20pt) .. node[below]{$b$}(b);
  \draw (a) .. controls +(+30:20pt) and +(+150:20pt) .. node[above]{$a$} (b);
  \draw (b) -- node[above] {$c$}(c);
\end{tikzpicture}
\end{center}

We note that when we describe a matroid in terms of a graph in this document we
refer to the matroid where the ground set are the edges of the graph and a set
is independent if the corresponding edges do not form a cycle. Call the rank of this matroid $r[M_{((ab)c)}]$. 
Therefore, the set of functions in $\G^3$ associated with the depicted
$\emptyset$-tree is given by:
$$\E^3 + r[U_3^1] \cdot \R_+ + r[U_3^2] \cdot \R_+ + r[M_{((ab)c)}] \cdot \R_+$$

Since all the $\emptyset$-trees are symmetric, then all the gross substitute
functions over $3$ elements are of the form:
$$\G^3 = \E^3 + r[U_3^1] \cdot \R_+ + r[U_3^2] \cdot \R_+ + \left(
r[M_{((ab)c)}] \cdot \R_+ \cup r[M_{((ac)b)}] \cdot \R_+ \cup r[M_{((bc)a)}]
\cdot \R_+ \right)$$

\subsubsection{Description of $\G^4$}  The description of $\G^4$ follows similarly as for $\G^3$, but with more cases. It is deferred to Appendix~\ref{sec:appG4}.

\subsection{Finding the counterexample}

The main idea  used to find counterexample $v$ is to exhibit  a specific tree
structure over $5$ elements that is complex enough so that, unlike for $n =
1,2,3,$ and $4$, node labels cannot be decomposed into binary valued vectors
which satisfy the integrability condition. We will try to mimic the same proof
used for $\G^3$ and $\G^4$ and find a set of trees for which the same proof
technique cannot be extended.
%Since matroids have second
%derivatives that are either $0$ or $1$, this indicates that a function $v$
%corresponding to such a tree signature and labels is not a positive linear
%combinations of matroids which are tree concordant with $v$.

%%%%%%%%%%%%%%%%%
\begin{figure}[h]
\centering
\begin{subfigure}{.33\textwidth}
  \centering
  \begin{tikzpicture}[scale=.8]
  \node[circle, draw=black, label=left:$\emptyset$:] {$m_1$}
    child {node[circle, draw=black] {$m_2$}
      child {node {$1,2,3$}}}
		child {node {$4,5$}};
\end{tikzpicture}
\end{subfigure}%
\begin{subfigure}{.33\textwidth}
  \centering
  \begin{tikzpicture}[scale=.8]
  \node[circle, draw=black, label=left:$\{5\}$:] {$m_3$}
    child {node[circle, draw=black] {$m_4$}
      child {node {$1,2$}}}
		child {node {$3,4$}};
\end{tikzpicture}
\end{subfigure}%
\begin{subfigure}{.33\textwidth}
  \centering
  \begin{tikzpicture}[scale=.8]
  \node[circle, draw=black, label=left:$\{1\}$:] {$m_5$}
    child {node[circle, draw=black] {$m_6$}
      child {node {$2,4,5$}}}
		child {node {$3$}};
\end{tikzpicture}
\end{subfigure}%
\caption{The key part of the tree structure for the counterexample.}
\label{fig:hard}
\end{figure}
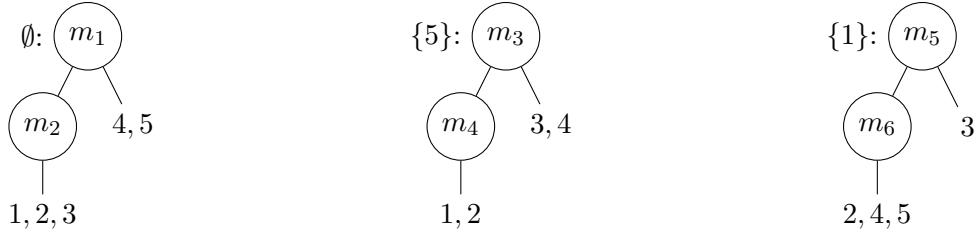

Consider the tree defined in Figure~\ref{fig:hard}. We can apply the
integrability conditions (\ref{eq:int}) to obtain the following relations among
the labels:
\begin{itemize}
\item $m_3 - m_5 = m_2 - m_1\geq 0$ from (\ref{eq:int}) with $S = \emptyset$ and $(i,j,k) = (1,3,5)$,
\item $m_6 - m_3 = m_1 - m_1 = 0$ from (\ref{eq:int}) with $S = \emptyset$ and $(i,j,k) = (1,4,5)$,
\item $m_4 - m_6 = m_2 - m_1 \geq 0$ from (\ref{eq:int}) with $S = \emptyset$
  and $(i,j,k) = (1,2,5)$.
\end{itemize}

Consider a function that satisfies that tree structure and has $m_2 - m_1 =
\Delta > 0$, then $m_4 = m_6 + \Delta = m_3 + \Delta = m_5 + 2 \Delta$.
Therefore, any parametrization of the space of functions sharing those trees
obtained by the same method as used for $\G^3$ and $\G^4$ will have a non-binary
coefficient and hence does not decompose the space in combinations of matroid
rank functions.

This is yet not a proof, since we do not know yet if there is GS functions with
that tree structure and if there is a parametrization obtained by other methods
that decomposes the space. Next, we complete this tree description and
(computationally) solve a linear program to find a function satisfying that tree
structure. Once we get a candidate functions that is the output of this program,
we write a second program that tries to write it as a convex combination of
matroid rank functions by explicitly enumerating over the set of all matroid
rank functions and creating one variable for each in the linear program.
Next, we verify that the program is infeasible and obtain a Farkas' type
certificate. Finally, since we do not want to rely on the correctness of the
enumeration and the computational steps, we give a human readable proof.

\section{Conclusion}
\label{sec:conclusion}

The class of gross substitutes is a well-studied family of valuation functions that has many different characterizations, but for which we do not know a constructive description. Our main result shows that gross substitutes cannot be constructed via positive linear combinations of matroid rank functions. We also give a new operation, called tree-concordant-sum, which provides a necessary and sufficient condition for the sum operation to preserve substitutability and which is used to find the counterexample for the main result.

In addition to affine transformations, strong quotient sum, and tree-concordant-sum, other operations are known to preserve substitutability. Two important examples are endowment or restriction \cite{HatfieldMilgrom} and convolution \cite{Murota96} or OR \cite{LehmannLehmannNisan}. It remains an important open question whether there is a collection of substitutability-preserving operations that allow constructing all gross substitutes from matroid rank functions, or another simple class of functions.

%Two of the particularly important
%ones are:

%\begin{itemize}
%  \item {\bf Endowment or restriction \cite{HatfieldMilgrom}.}
%    Given a valuation function defined on
%    $n$ elements $v:2^{N} \rightarrow \R$ and a subset $S \subseteq N$, we can
%    define a function defined on $N \setminus S$ elements as $w:2^{N\setminus S}
%    \rightarrow \R$ such that $w(T) = v(T \vert S)$.
%
%  \item {\bf Convolution \cite{Murota96} or OR \cite{LehmannLehmannNisan}.}
%    Given two valuation functions $v_1, v_2 : 2^N \rightarrow \R$ we define the
%    convolution $\tilde{v} = v_1 * v_2$ as the function $\tilde{v} : 2^N
%    \rightarrow \R$ such that $\tilde{S} = \max_{T \subseteq S} v_1(T) + v_2(S
%    \setminus T)$.
%\end{itemize}
%
%The main question left open is whether those operations, together with the newly
%added operation of tree-concordant-sum is enough to construct all GS functions.

\section*{Acknowledgements}
We thank Kazuo Murota and Akiyoshi Shioura for their comments on an earlier
version of this manuscript and their various suggestions to improve the
presentation. We would also like to thank the mantainers of the database of
matroids \cite{databasematroid} which was in invaluable resource in this
project.

Eric is supported by a Google PhD Fellowship.

\bibliographystyle{plainnat}

\bibliography{sigproc}

\newpage
\section*{Appendix}
\appendix
% !TeX root = main.tex

\section{Missing Proofs from Section~\ref{sec:prelim}}

\begin{theorem}\label{thm:matroid_submodular}
$\MM^n = \{v \in \G^n; v(\emptyset) = 0; \partial_i v(S) \in \{0,1\}, \forall S
\subseteq [n] \} $.
\end{theorem}

It follows from the following characterization of matroid rank functions that
can be found in Section 39.7 of Schrijver \cite{schrijver2003combinatorial}
(rephrased in the language of discrete derivatives):

\begin{lemma} Let $r$ be a valuation function, then  $r \in \MM^n$ if and only if  $r(\emptyset) = 0$,
  $\partial_i r(S) \in \{0,1\}$ and $\partial_{ij} r(S) \leq 0$.
\end{lemma}

This means that matroid rank functions are exactly the submodular functions that
have $\{0,1\}$-marginals. Using this lemma, we can now prove
Theorem~\ref{thm:matroid_submodular}:

\proof{Proof of Theorem~\ref{thm:matroid_submodular}}
  The inclusion $\supseteq$ follows directly from the previous lemma and the
  fact that every GS function is submodular.

  For the inclusion $\MM^n \subseteq \G^n$, we first note that if $r \in \MM^n$,
  then $\partial_{ij}r(S) = \partial_i r(S\cup j) - \partial_i r(S) \in
  \{0,-1\}$ since the first derivatives are in $\{0,1\}$ and the second
  derivatives are non-positive. Moreover, $\partial_{ij}r(S) = -1$ if and only if
  $\partial_i r(S\cup j)  = \partial_j r(S\cup i) = 0$ and $\partial_i r(S) = 
  \partial_j r(S) = 1$. Therefore the condition:
  $$\partial_{ij} v(S) \leq \max[ \partial_{ik} v(S), \partial_{kj} v(S)  ]$$
  is violated only when $\partial_{ij} v(S) = 0$ and $\partial_{ik} v(S) =
  \partial_{kj} v(S) = -1$. This implies that $\partial_i v(S) = \partial_j v(S)
  = \partial_k v(S) = 1$. But since $\partial_{ij} v(S) = 0$ we must have
  $v(S \cup ijk) \geq v(S\cup ij) = v(S) + 2$. This implies
\revision{ $\partial_{ij}v(S\cup k) > 0$ which contradicts submodularity.}  
\endproof

\section{Missing Proofs from Section~\ref{sec:building_block}}\label{appendix:submodular_constr}

The main ingredient in Theorem \ref{thm:submodular_construction} is the
following characterization of monotone integer valued submodular functions:

\begin{lemma}[\citet{schrijver2003combinatorial}, section 44.6b]
  Every monotone integer valued submodular function that evaluates to zero at
  the empty set can be obtained from a
  matroid rank function by item grouping. Formally: if $v:2^{[n]} \rightarrow
  \Z$ is a monotone submodular function (where monotone means that $\partial_i
  v(S) \geq 0$) with $v(\emptyset) = 0$, then there is a matroid $\M$ defined on
  a set $U$ and a partition $(X_1, \hdots, X_n)$ of $U$ such that $v(S) =
  r_\M(\cup_{s \in S} X_s)$.
\end{lemma}

%{\color{blue} [I am til unsure whether to have this section or just move it to
%the appendix. I wanted to give a simple example of contructive description. I am
%afraid with this paper to have too many digressions. On the other hand, I think
%they are somehow useful. Maybe we can move some of that to the appendix if we
%feel that the paper is too long or not enough to the point.]}

Therefore we only need to show that all submodular functions can be constructed
from monotone integer valued submodular functions using positive linear
combinations and affine transformations. To show this we start by viewing each
submodular function as a vector in $\R^{2^n}$ indexed by the subsets of $[n]$.
From this perspective, the set of submodular functions $\SS^n$ 
correspond to the set of
$\R^{2^n}$-points satisfying the linear inequalities given by $\partial_{ij}
v(S) \leq 0$, which is a system of homogeneous linear inequalities. The set of
solutions of such system is usually called a 
\emph{polyhedral cone}. A classic result in convex analysis (see
\cite{rockafellar2015convex} for example) says that every polyhedral cone is finitely generated,
i.e., every point can be written as a positive combination of a finite set of
points. In other words, there is a finite set $v_1, \hdots, v_k \in \SS^n$ such
that:
$$\SS^n = \left\{ \textstyle \sum_{i=1}^k \alpha_i v_i ; \alpha_i \geq 0 \right\} $$
When this set is minimal, those are called extremal rays of the cone. Also from
convex analysis, if the constraints have rational coefficients then there is a
set of extremal rays with integer coefficients.

Finally, observe that we can construct general integer value submodular
functions from monotone ones by applying an affine transformation. For each
$v_i$, let $M = \min_{S
\subseteq [n], j \notin S} \partial_j v_i(S)$ and then define a normalized
version of $v_i$ as:
$$\bar v_i(S) = v_i(S) - M \abs{S} - v_i(\emptyset)$$
It is simple to see that $\bar v_i(S)$ is monotone and evaluates to zero at the
empty set and that $v_i$ can be constructed from $\bar v_i$ using an affine
transformation.

\paragraph{Extremal submodular functions} One can ask whether the \emph{item
grouping} operation is necessary. Equivalently, are all the extremal rays
matroid rank functions? For $\SS^2$ and $\SS^3$ all submodular functions are
convex combinations of matroid rank functions (modulo affine transformations).
For $n=4$, however, the following submodular function is extremal and is not a
matroid rank function: define $f(S)$ over $\{a,b,c,d\}$ such that
$f(S) = 0$ for $\abs{S} \leq 1$, $f(ab) = f(bd) = \revision{f(bc)} = f(acd) = -1$,
$f(ac) = f(ad) = f(cd) = 0$, $f(abc) = f(abd) = f(bcd) = -2$ and $f(abcd) = -3$.

In general the set of extremal submodular functions can be obtained using the
standard technique of converting between the $H$-representation and
$V$-representation of a cone. See \citet{ziegler2012lectures} for a
complete discussion and the LRS package \cite{lrs} for an implementation of such
algorithms.

\section{Missing Proofs from Section~\ref{sec:counterexample}}
\label{sec:appcounterexample}

\subsection{Checking $\G^5_0$-conditions for candidate}\label{sec:checkcond}

Below we check conditions $\partial_{ij}v(S) \leq \max[\partial_{ik}v(S),
\partial_{kj}v(S)] \leq 0$ for the candidate function in
Table~\ref{tab:function}. There are $40$ inequalities to be checked, which we do below.

\begin{tiny}
\noindent \begin{minipage}[t]{0.5\textwidth}
$$\begin{aligned}
&-1 = \partial_{3,2}v(\emptyset) \leq \max(\partial_{3,1}v(\emptyset),\partial_{1,2}v(\emptyset)) = \max(-1,-1) = -1 \leq 0\\
&0 = \partial_{4,2}v(\emptyset) \leq \max(\partial_{4,1}v(\emptyset),\partial_{1,2}v(\emptyset)) = \max(0,-1) = 0 \leq 0\\
&0 = \partial_{4,3}v(\emptyset) \leq \max(\partial_{4,1}v(\emptyset),\partial_{1,3}v(\emptyset)) = \max(0,-1) = 0 \leq 0\\
&0 = \partial_{4,3}v(\emptyset) \leq \max(\partial_{4,2}v(\emptyset),\partial_{2,3}v(\emptyset)) = \max(0,-1) = 0 \leq 0\\
&0 = \partial_{5,2}v(\emptyset) \leq \max(\partial_{5,1}v(\emptyset),\partial_{1,2}v(\emptyset)) = \max(0,-1) = 0 \leq 0\\
&0 = \partial_{5,3}v(\emptyset) \leq \max(\partial_{5,1}v(\emptyset),\partial_{1,3}v(\emptyset)) = \max(0,-1) = 0 \leq 0\\
&0 = \partial_{5,3}v(\emptyset) \leq \max(\partial_{5,2}v(\emptyset),\partial_{2,3}v(\emptyset)) = \max(0,-1) = 0 \leq 0\\
&0 = \partial_{5,4}v(\emptyset) \leq \max(\partial_{5,1}v(\emptyset),\partial_{1,4}v(\emptyset)) = \max(0,0) = 0 \leq 0\\
&0 = \partial_{5,4}v(\emptyset) \leq \max(\partial_{5,2}v(\emptyset),\partial_{2,4}v(\emptyset)) = \max(0,0) = 0 \leq 0\\
&0 = \partial_{5,4}v(\emptyset) \leq \max(\partial_{5,3}v(\emptyset),\partial_{3,4}v(\emptyset)) = \max(0,0) = 0 \leq 0\\
&0 = \partial_{4,3}v(1) \leq \max(\partial_{4,2}v(1),\partial_{2,3}v(1)) = \max(-1,0) = 0 \leq 0\\
&0 = \partial_{5,3}v(1) \leq \max(\partial_{5,2}v(1),\partial_{2,3}v(1)) = \max(-1,0) = 0 \leq 0\\
&-1 = \partial_{5,4}v(1) \leq \max(\partial_{5,2}v(1),\partial_{2,4}v(1)) = \max(-1,-1) = -1 \leq 0\\
&-1 = \partial_{5,4}v(1) \leq \max(\partial_{5,3}v(1),\partial_{3,4}v(1)) = \max(0,0) = 0 \leq 0\\
&0 = \partial_{4,3}v(2) \leq \max(\partial_{4,1}v(2),\partial_{1,3}v(2)) = \max(-1,0) = 0 \leq 0\\
&0 = \partial_{5,3}v(2) \leq \max(\partial_{5,1}v(2),\partial_{1,3}v(2)) = \max(-1,0) = 0 \leq 0\\
&-1 = \partial_{5,4}v(2) \leq \max(\partial_{5,1}v(2),\partial_{1,4}v(2)) = \max(-1,-1) = -1 \leq 0\\
&-1 = \partial_{5,4}v(2) \leq \max(\partial_{5,3}v(2),\partial_{3,4}v(2)) = \max(0,0) = 0 \leq 0\\
&0 = \partial_{5,4}v(1,2) \leq \max(\partial_{5,3}v(1,2),\partial_{3,4}v(1,2)) = \max(0,0) = 0 \leq 0\\
&0 = \partial_{4,2}v(3) \leq \max(\partial_{4,1}v(3),\partial_{1,2}v(3)) = \max(0,0) = 0 \leq 0\\
\end{aligned}$$
\end{minipage}\begin{minipage}[t]{0.5\textwidth}
$$\begin{aligned}
&0 = \partial_{5,2}v(3) \leq \max(\partial_{5,1}v(3),\partial_{1,2}v(3)) = \max(0,0) = 0 \leq 0\\
&-1 = \partial_{5,4}v(3) \leq \max(\partial_{5,1}v(3),\partial_{1,4}v(3)) = \max(0,0) = 0 \leq 0\\
&-1 = \partial_{5,4}v(3) \leq \max(\partial_{5,2}v(3),\partial_{2,4}v(3)) = \max(0,0) = 0 \leq 0\\
&-1 = \partial_{5,4}v(1,3) \leq \max(\partial_{5,2}v(1,3),\partial_{2,4}v(1,3)) = \max(-1,-1) = -1 \leq 0\\
&-1 = \partial_{5,4}v(2,3) \leq \max(\partial_{5,1}v(2,3),\partial_{1,4}v(2,3)) = \max(-1,-1) = -1 \leq 0\\
&-1 = \partial_{3,2}v(4) \leq \max(\partial_{3,1}v(4),\partial_{1,2}v(4)) = \max(-1,-2) = -1 \leq 0\\
&-1 = \partial_{5,2}v(4) \leq \max(\partial_{5,1}v(4),\partial_{1,2}v(4)) = \max(-1,-2) = -1 \leq 0\\
&-1 = \partial_{5,3}v(4) \leq \max(\partial_{5,1}v(4),\partial_{1,3}v(4)) = \max(-1,-1) = -1 \leq 0\\
&-1 = \partial_{5,3}v(4) \leq \max(\partial_{5,2}v(4),\partial_{2,3}v(4)) = \max(-1,-1) = -1 \leq 0\\
&0 = \partial_{5,3}v(1,4) \leq \max(\partial_{5,2}v(1,4),\partial_{2,3}v(1,4)) = \max(0,0) = 0 \leq 0\\
&0 = \partial_{5,3}v(2,4) \leq \max(\partial_{5,1}v(2,4),\partial_{1,3}v(2,4)) = \max(0,0) = 0 \leq 0\\
&0 = \partial_{5,2}v(3,4) \leq \max(\partial_{5,1}v(3,4),\partial_{1,2}v(3,4)) = \max(0,-1) = 0 \leq 0\\
&-1 = \partial_{3,2}v(5) \leq \max(\partial_{3,1}v(5),\partial_{1,2}v(5)) = \max(-1,-2) = -1 \leq 0\\
&-1 = \partial_{4,2}v(5) \leq \max(\partial_{4,1}v(5),\partial_{1,2}v(5)) = \max(-1,-2) = -1 \leq 0\\
&-1 = \partial_{4,3}v(5) \leq \max(\partial_{4,1}v(5),\partial_{1,3}v(5)) = \max(-1,-1) = -1 \leq 0\\
&-1 = \partial_{4,3}v(5) \leq \max(\partial_{4,2}v(5),\partial_{2,3}v(5)) = \max(-1,-1) = -1 \leq 0\\
&0 = \partial_{4,3}v(1,5) \leq \max(\partial_{4,2}v(1,5),\partial_{2,3}v(1,5)) = \max(0,0) = 0 \leq 0\\
&0 = \partial_{4,3}v(2,5) \leq \max(\partial_{4,1}v(2,5),\partial_{1,3}v(2,5)) = \max(0,0) = 0 \leq 0\\
&0 = \partial_{4,2}v(3,5) \leq \max(\partial_{4,1}v(3,5),\partial_{1,2}v(3,5)) = \max(0,-1) = 0 \leq 0\\
&0 = \partial_{3,2}v(4,5) \leq \max(\partial_{3,1}v(4,5),\partial_{1,2}v(4,5)) = \max(0,-1) = 0 \leq 0\\
\end{aligned}$$
\end{minipage}
\end{tiny}

\subsection{Weighted matroids}

\begin{lemma}[\citet{shioura2012matroid}]
\label{lem:weighted}
Any weighted matroid rank functions can be written as a positive linear combination of unweighted matroid rank functions.
\end{lemma}
\proof{Proof.}
 Let $w_1 \geq \ldots \geq w_n \geq 0$ be the weights of the $n$ elements $[n] := \{1,
  \ldots, n\}$ of a weighted matroid rank function $v$ associated to matroid
  $\M$. Let $\M_i$ be the matroid $\M$ restricted to  elements $[i]$ and $r_i$ be
  the unweighted matroid rank function over  elements $[i]$ associated with
  matroid $\M_i$. Since the greedy algorithm finds a maximum weight base of a
  weighted matroid, we have \begin{align*}
v(S) & = \sum_{i \in [n]} w_i \left( r(S \cap [i]) - r(S \cap [i-1]) \right) \\
& = \sum_{i \in [n]} \sum_{j = i}^{n} (w_j - w_{j+1}) \left( r(S \cap [i]) - r(S \cap [i-1]) \right) \\
& = \sum_{ j \in [n]} (w_j - w_{j+1})  \cdot \revision{r(S \cap \{j\})} \\
& = \sum_{j \in [n]} (w_j - w_{j+1})  \cdot \revision{r_j(S)}.  
\end{align*} 
\endproof

\section{Integrability Conditions}\label{appendix:integrability}

Given differentiable functions $b_i : \R^n \rightarrow \R$ for $i = 1, \ldots,
n$, it is well known that there is a function $f$ such that $b_i(x) = \partial f(x) /
\partial x_i$ if and only if the functions $b_i$ satisfy the conditions $\partial b_i /
\partial x_j = \partial b_j / \partial x_i$. \revision{In physics, those
correspond to the necessary and sufficient conditions for a field to be a
conservative field. We refer to Section 10.16 of \cite{apostol1969calculus}
for a complete discussion. Those conditions can also be derived as a special
case of Stoke's Theorem.} The exact same condition provides
integrability over the hypercube:

\begin{lemma}\label{lemma:int:1}
  Given functions $\beta_i : 2^{[n] \setminus i} \rightarrow \R$ for
  $i \in [n]$, then there exists a function $v$ such that $\beta_i = \partial_i v$
  for all $i$ if and only if $\partial_i \beta_j = \partial_j \beta_i$ for all $i \neq j$.
\end{lemma}

\proof{Proof.}
  Given $\beta_i$ satisfying $\partial_i \beta_j = \partial_j \beta_i$, fix any
  order among the elements in $[n]$ and let $S_{<i} = \{j \in S; j < i\}$. Now,
  define $v$ as follows:
  $$v(S) = \sum_{i \in S} \beta_i(S_{<i})$$
  First we argue that the definition is order independent, i.e., for any
  ordering of the elements, we construct the same function $v$. To see that,
  start for an arbitrary order and swap a pair of adjacent elements $i<j$. Then
  if only one is in $S$, this doesn't change $v(S)$. If both are in $S$, we
  change the definition of $v(S)$ from:
  $$ \hdots + \beta_i( T ) + \beta_j (T\cup i) + \hdots$$
  to the following (where the terms in $\hdots$ are left unchanged):
  $$ \hdots + \beta_j( T ) + \beta_i (T\cup j) + \hdots$$
  Since $\partial_j \beta_i(T) = \partial_i \beta_j(T)$ we have $\beta_i( T ) +
  \beta_j (T\cup i)  = \beta_j( T ) + \beta_i (T\cup j)$. This is equivalent to
  the notion of \emph{path-independence} for continuous functions \revision{(see Section
  10.17 of \cite{apostol1969calculus})}.

  Now, fixed $j$, we can assume without loss of generality that $j$ is placed
  in the end of the ordering. Therefore by the definition of $v$ we have $v(S
  \cup j) = v(S) + \beta_j(S)$ so $\beta_j(S) = \partial_j v(S)$.  
\endproof

We can now obtain second order integrability conditions from the first order
ones easily:

\begin{lemma}\label{lemma:int:2}
  Given functions $\alpha_{ij} : 2^{[n] \setminus ij} \rightarrow \R$ for
  $i \neq j$, then there exists a function $v$ such that $\alpha_{ij} =
  \partial_{ij} v$ for all $i,j$ if and only if $\alpha_{ij} = \alpha_{ji}$ and
  $$\partial_i \alpha_{jk} = \partial_j \alpha_{ik} = \partial_k \alpha_{ij}
  \quad \forall \text{ distinct } i,j,k.$$
\end{lemma}

\proof{Proof.}
  First observe that by first order integrability conditions (Lemma
  \ref{lemma:int:1}) we can find for each $i$, a function $\beta_i$ such that
  $\alpha_{ij} = \partial_j \beta_i$ since $\partial_k \alpha_{ij} = \partial_j
  \alpha_{ik}$. Now observe that the functions $\beta_1, \hdots, \beta_n$
  constructed satisfy first order integrability conditions, since $\partial_j
  \beta_i = \alpha_{ij} = \partial_i \beta_j$ so there is $v$ such that $\beta_i
  = \partial_i v$.  
\endproof

\begin{corollary}\label{cor:int:3}
  Given functions  $\alpha_{ij} : 2^{[n] \setminus ij}
  \rightarrow \R$ satisfying integrability conditions in the previous lemma, all
  $v : 2^{[n]} \rightarrow \R$ are affine transformations of the function:
  $$v(S) = \sum_{i<j; i,j \in S} \alpha_{ij}(S_{<i})$$
\end{corollary}

\proof{Proof.}
  Using Lemma \ref{lemma:int:1} we can reconstruct the functions $\beta_i$ as:
  $$\beta_i(S) = \beta_i(\emptyset) + \sum_{j \in S} \alpha_{ij}(S_{<j})$$
  Applying the same process for reconstructing $v$ from $\beta_i$, we get:
  $$\begin{aligned}
    v(S) & = v(\emptyset) + \sum_{i \in S} \beta_i(S_{<i}) =  v(\emptyset) + 
  \sum_{i \in S} \left[ \beta_i(\emptyset) +  \sum_{j \in S_{<i}}
    \alpha_{ij}(S_{<j}) \right] \\ & = v(\emptyset) + \sum_{i \in S}
    \beta_i(\emptyset) + \sum_{j < i; i,j \in S}  \alpha_{ij}(S_{<j})   
  \end{aligned}$$
\endproof

\section{Description of $\G^4$}
\label{sec:appG4}

For the case of $\G^4$, we start by observing that up to renaming
the items, the $\emptyset$-tree must have one of two forms, which we will refer
as the shallow tree and the deep tree respectively.

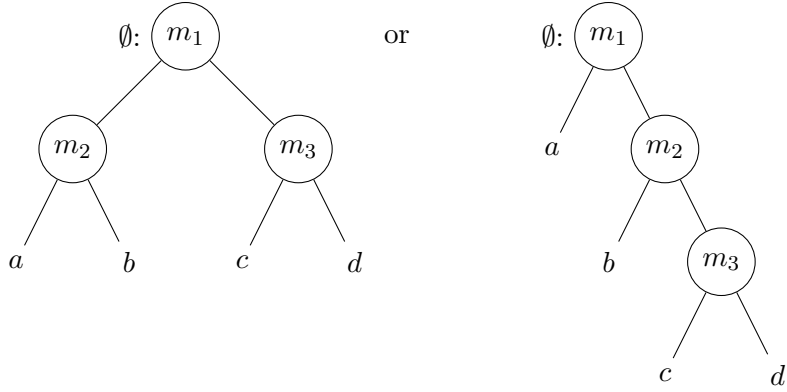
\begin{center}
  \begin{tikzpicture}[level/.style={sibling distance=30mm/#1}]
  \node[circle, draw=black, label=left:$\emptyset$:] {$m_1$}
    child {node[circle, draw=black] {$m_2$}
      child {node {$a$}}
      child {node {$b$}}}
    child {node[circle, draw=black] {$m_3$}
      child {node {$c$}}
      child {node {$d$}}};
  \node[xshift=80pt]{or};
	\node[xshift=160pt, circle, draw=black, label=left:$\emptyset$:]{$m_1$}
    child [sibling distance = 15mm]{node {$a$}}
    child [sibling distance = 15mm]{node[circle, draw=black] {$m_2$}
      child {node {$b$}}
      child {node[circle, draw=black] {$m_3$}
        child [sibling distance = 15mm] {node {$c$}}
        child [sibling distance = 15mm] {node {$d$}}}};
\end{tikzpicture}
\end{center}

\paragraph{Shallow Tree Case}

Assume we have a valuation function $v \in \G^4_0$ whose $\emptyset$-tree is shallow (i.e.,
is like the left diagram above). We know $m_1 \leq m_2$ and $m_1 \leq m_3$. For convenience of notation we will
refer to $m_1 = x$, $m_2 = x + y$ and $m_3 = x + z$ for $x,y,z \geq 0$. Applying Lemma~\ref{lem:welldefined} with $S = \emptyset$ and for every triple of elements, we get:

$$\begin{aligned}
w_1 =: \Delta_{ac}^d = \Delta_{ad}^c = \Delta_{cd}^a - z\\
w_2 =: \Delta_{bc}^d = \Delta_{bd}^c = \Delta_{cd}^b - z\\
  w_3 =: \Delta_{ac}^b = \Delta_{bc}^a = \Delta_{ab}^c - y\\
  w_4 =: \Delta_{ad}^b = \Delta_{bd}^a = \Delta_{ab}^d - y
\end{aligned}
$$

If $w_3 \neq w_4$ we can wlog (up to permuting the identity of the items) assume
that $w_3 < w_4$. In such case, observe that:
$\Delta_{ac}^b = w_3$, $\Delta_{ad}^b = w_4$, and $\Delta_{cd}^b = w_2 + z$. Since the minimum value is repeated among $\Delta_{ac}^b$, $\Delta_{ad}^b$ and
$\Delta_{cd}^b$ we must have $w_3 = w_2 + z$. Now, looking at the substitution symbols for the $\{a\}$-tree, we get:
$\Delta_{bc}^a = w_3$, $\Delta_{bd}^a = w_4$, and $\Delta_{cd}^a = w_1 + z$;
so we must have by the same argument:
$w_3 = w_1 + z$. In particular we will get:
$w_1 = w_2 = w_3 - z = w_4 - z - \revisiontwo{q}$ for some $\revisiontwo{q}\geq 0$. This means that we can write:
\begin{equation}\label{eq:shallow_case}\tag{S}
w_1 = w \qquad w_2 = w \qquad w_3 = w + z \qquad w_4 = w + z + \revisiontwo{q}.
\end{equation}

\revision{If $w_3 = w_4$, then there are two cases.  If $w_1 \neq w_2,$ then up
to permuting the identity of items this case is identical to $w_3 \neq w_4$ and
we can obtain \eqref{eq:shallow_case} with indices permuted. If $w_1 = w_2$ and $w_3 = w_4,$ \revisiontwo{then up
to permuting the identity of items this case is a special case of \eqref{eq:shallow_case} with $ \revisiontwo{q}= 0$.}
Therefore it is w.l.o.g. to focus on the setup in \eqref{eq:shallow_case}.

The values of $w_1$, $w_2$, $w_3$ and $w_4$ define} the $\{i\}$-trees for all
$i = a,b,c,d$. It is possible now to reconstruct $v(S)$ for $S \neq \emptyset$
with Corollary \ref{cor:integration}:
$$\begin{aligned}
  &v(abc) = -\Delta_{ab}^\emptyset -\Delta_{ac}^\emptyset - \Delta_{bc}^a = -2x
  -  y - w - z\\
  &v(abd) = -\Delta_{ab}^\emptyset -\Delta_{ad}^\emptyset - \Delta_{bd}^a = -2x -
  y - w - z - \revisiontwo{q}\\
  &v(acd) = -\Delta_{ac}^\emptyset -\Delta_{ad}^\emptyset - \Delta_{cd}^a = -2x -
  w - z\\
  &v(bcd) = -\Delta_{bc}^\emptyset -\Delta_{bd}^\emptyset - \Delta_{cd}^b = -2x
  - z - w \\
  & v(abcd) = -3x - y - 2z - 2w - \revisiontwo{q}- t
\end{aligned}$$
for some $t \geq 0$. This gives us a valuation $v$ in gross substitutes parametrized by
$(x,y,z,w,u,t) \in \R^6_+$. Therefore, every valuation $v$ in $\G^4_0$ whose
$\emptyset$-tree is shallow can be written as a non-negative combination of $6$
\emph{extremal} valuation functions. The extremal valuations are obtained when
we set one of the coefficients to one and all coefficients to zero. It is
instructive to see which valuations are those.

\begin{itemize}
  \item $x = 1$ and all other coefficients are zero. We obtain a valuation function
    such that $v(S) = 0$ for $\abs{S} \leq 1$, $v(S) = -1$ for $\abs{S} = 2$,
    $v(S) = -2$ for $\abs{S} = 3$ and $v(S) = -3$ for $\abs{S} = 4$. We get  $\tilde{v}(S) = 0$ for $S
    = \emptyset$ and $\tilde{v}(S) = 1$ for $S \neq \emptyset$, which is $r[U^1_4]$.
  \item $y=1$ and all other coefficients are zero. Then $v(S) = -1$ if $\{a,b\}
    \subseteq S$ and $v(S) = 0$ otherwise, and $\tilde{v}$ is the rank
    function of the following matroid:
    
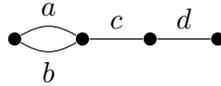
\begin{center}
      \begin{tikzpicture}[baseline=-0.65ex, scale=.6]
      \tikzstyle{vertex}=[circle,fill=black,minimum size=5pt,inner sep=0pt];
      \node[vertex] (a) at (0,0) {};
      \node[vertex] (b) at (1.5,0)  {};
      \node[vertex] (c) at (3,0) {};
      \node[vertex] (d) at (4.5,0) {};
      \draw (a) .. controls +(-30:20pt) and +(-150:20pt) .. node[below]{$b$}(b);
      \draw (a) .. controls +(+30:20pt) and +(+150:20pt) .. node[above]{$a$} (b);
      \draw (b) -- node[above] {$c$}(c);
      \draw (c) -- node[above] {$d$}(d);
      \end{tikzpicture}
    \end{center}
  \item $z = 1$ and all other coefficients are zero. Then $v(S) = 0$ for
    $\abs{S} \leq 1$, $v(cd) = -1$, $v(S) = 0$ for all other $S$ with $\abs{S} =
    2$,  $v(S) = -1$ for all  $S$ with $\abs{S} = 3$, and
    finally $v(abcd) = -2$. So $\tilde{v}$ is  the rank function of the matroid:
    
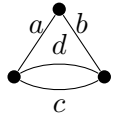
\begin{center}
      \begin{tikzpicture}[baseline=1ex, scale=.6]
        \tikzstyle{vertex}=[circle,fill=black,minimum size=5pt,inner sep=0pt];
        \node[vertex] (a) at (0,0) {};
        \node[vertex] (b) at (2,0)  {};
        \node[vertex] (c) at (1,1.5) {};
        \draw (a) .. controls +(-30:20pt) and +(-150:20pt) .. node[below]{$c$}(b);
        \draw (a) .. controls +(+30:20pt) and +(+150:20pt) .. node[above]{$d$} (b);
        \draw (a) -- node[above] {$a$}(c);
      \draw (b) -- node[above] {$b$}(c); \end{tikzpicture}
    \end{center}
  \item $w = 1$ and all other coefficients are zero, then $v(S) = 0$ for
    $\abs{S} \leq 2$, $v(S) = -1$ for $\abs{S} = 3$ and $v(S) = -2$ for $\abs{S}
    = 4$, and $\tilde{v}$ is  $r[U_4^2]$.
  \item $\revisiontwo{q}= 1$ and all other coefficients are zero, then $v(S) = -1$ if
    $\{a,b,d\} \subseteq S$ and $v(S) = 0$ otherwise. So,  $\tilde{v}(S) = v(S)
    +  \abs{S}$ is the rank function of the matroid:
    
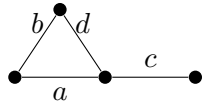
\begin{center}
      \begin{tikzpicture}[baseline=1ex, scale=.6]
        \tikzstyle{vertex}=[circle,fill=black,minimum size=5pt,inner sep=0pt];
        \node[vertex] (a) at (0,0) {};
        \node[vertex] (b) at (2,0)  {};
        \node[vertex] (c) at (1,1.5) {};
        \node[vertex] (d) at (4,0) {};
        \draw (a) -- node[below] {$a$}(b);
        \draw (a) -- node[above] {$b$}(c);
      \draw (b) -- node[above] {$d$}(c);
      \draw (b) -- node[above] {$c$}(d); \end{tikzpicture}
    \end{center}
  \item if $t = 1$ and all other coefficients are zero, then $v(S) = 0$ if
    $\abs{S} \leq 3$ and $v(S) = -1$ for $\abs{S} = 4$, therefore,
    $\tilde{v}(S)$ is 
    $r[U_4^3]$.
\end{itemize}

Therefore we identified the following cone which is a subset of $\G^4$:
$$\E^4 + \sum_{j=1}^6 r_j \cdot \R_+  $$
where $r_1, \hdots, r_6$ are the rank functions of the matroids identified in
the previous items. Also, since $\G^4$ is symmetric with respect to permutations
of the identities of the items, we can obtain $11$ other cones by permuting the
identities of the items.

\paragraph{Deep Tree Case.}

Assume now that we have a valuation $v \in \G^4_0$ whose $\emptyset$-tree is
deep. We know $m_1 \leq m_2 \leq m_3$, so we will refer for convenience to $m_1
= x$, $m_2 = x+y$ and $m_3 = x+y+z$ for $x,y,z \geq 0$. By applying again
Lemma \ref{lem:welldefined} we obtain:

$$\begin{aligned}
w_1 =: \Delta_{ab}^c = \Delta_{ac}^b = \Delta_{bc}^a - y\\
w_2 =: \Delta_{ab}^d = \Delta_{ad}^b = \Delta_{bd}^a - y\\
  w_3 =: \Delta_{ac}^d = \Delta_{ad}^c = \Delta_{cd}^a - y - z\\
  w_4 =: \Delta_{bc}^d = \Delta_{bd}^c = \Delta_{cd}^b - z
\end{aligned}
$$

It is instructive to re-write the substitution symbols to write together the
symbols for the same tree:

$$
\begin{aligned}
& a : \\
& \Delta_{bc}^a = w_1 + y \\
& \Delta_{bd}^a = w_2 + y \\
& \Delta_{cd}^a = w_3 + y + z
\end{aligned}
\qquad
\begin{aligned}
& b : \\
& \Delta_{ac}^b = w_1 \\
& \Delta_{ad}^b = w_2 \\
& \Delta_{cd}^b = w_4 + z
\end{aligned}
\qquad
\begin{aligned}
& c : \\
& \Delta_{ab}^c = w_1 \\
& \Delta_{ad}^c = w_3 \\
& \Delta_{bd}^c = w_4
\end{aligned}
\qquad
\begin{aligned}
& d : \\
& \Delta_{ab}^d = w_2 \\
& \Delta_{ac}^d = w_3 \\
& \Delta_{bc}^d = w_4
\end{aligned}
$$

In each column, at least two values are equal and the third value is at least as
large as the two equal values. Let's look at the symbols for the $\{c\}$-tree.
There are four possibilities:

\begin{enumerate}
\item $w_1 = w_3 < w_4$. In such case, by looking at the $\{d\}$-tree, since
$w_3 < w_4$, it must be the case that $w_2 = w_3$ since the minimum substitution
symbol in the $\{d\}$-tree must be repeated. Therefore we must have $w_1 = w_2 = w_3 = w$ and $w_4 = w + \revisiontwo{q}$. It is simple to see that if this condition is true, for every $\{i\}$-tree,
the minimum value is repeated. 
\item $w_1 = w_4 < w_3$. In such case, by looking at the $\{d\}$-tree, since
\revision{$w_4 < w_3$,} it must be the case that $w_2 = w_4$ since the minimum substitution
symbol in the $\{d\}$-tree must be repeated. Therefore we must have $w_1 = w_2 = w_4 = w$ and $w_3 = w + \revisiontwo{q}$.
\item $w_3 = w_4 < w_1$. Now, we consider other three possibilities for the
$\{b\}$-tree:
\begin{enumerate}
\item $w_1 = w_4 + z \leq w_2$, so we have $w_1 = w + z$, $w_2 = w + z + \revisiontwo{q}$ and $w_3 = w_4 = w$,
\item $w_2 = w_4 + z < w_1$, so we have:
$w_1 = w+z+u$, $w_2 = w + z$, and $w_3 = w_4 = w$.
\item $w_1 = w_2 < w_4 + z$, so we have $ w_1 = w + \revisiontwo{q}$, $w_2 = w + \revisiontwo{q}$,  $w_3 = w_4 = w$, and $ z = \revisiontwo{q}+ \revisiontwo{s} $.
\end{enumerate}
\item $w_1 = w_3 = w_4$. By inspecting the $\{b\}$-tree, we must have $w_2 =
w_1$. Since this is a special case of the previous cases, we ignore this case
from now on.\\
\end{enumerate}

We note that in either case, we have:

$$\begin{aligned}
  &v(abc) = -\Delta_{ab}^\emptyset -\Delta_{ac}^\emptyset - \Delta_{bc}^a = -2x
  - y - w_1\\
  &v(abd) = -\Delta_{ab}^\emptyset -\Delta_{ad}^\emptyset - \Delta_{bd}^a = -2x -
  y - w_2 \\
  &v(acd) = -\Delta_{ac}^\emptyset -\Delta_{ad}^\emptyset - \Delta_{cd}^a = -2x -
  y - z - w_3\\
  &v(bcd) = -\Delta_{bc}^\emptyset -\Delta_{bd}^\emptyset - \Delta_{cd}^b = -2x
  - 2y - z - w_4
\end{aligned}$$

\revisiontwo{By Corollary~\ref{cor:integration}, we have $$v(abcd) =
-\Delta^{\emptyset}_{ab} - \Delta^{\emptyset}_{ac} - \Delta^{\emptyset}_{ad} -
\Delta_{bc}^a - \Delta_{bd}^a - \Delta_{cd}^{ab} = −3x − 2y − w_1 − w_2 − \Delta_{cd}^{ab}.$$
Next, we get
$$\Delta^{ab}_{cd} = \begin{cases} t + \revisiontwo{q}+ z & \text{case 1 and case 2} \\ t & \text{case 3a and case 3b} \\ t + \revisiontwo{s} & \text{case 3c} \end{cases}$$
for some $t \geq 0$ where case 1 is by
Lemma~\ref{lem:welldefined} with $S = \{b\}$ and triplet $acd$, case 2  and case 3c  are by
Lemma~\ref{lem:welldefined} with  $S = \{a\}$ and triplet $bcd$, and case 3a and case 3b are since we simply have $\Delta^{ab}_{cd} \geq 0$. We obtain 
$$v(abcd) = \begin{cases} −3x − 2y − 2w  - z - \revisiontwo{q}- t & \text{case 1 and case 2} \\ −3x − 2y − 2w  -2z -\revisiontwo{q}- t & \text{case 3a and case 3b} \\ −3x − 2y − 2w - 2\revisiontwo{q}  - \revisiontwo{s} - t& \text{case 3c} \end{cases}$$}

Now, following the same procedure used in the previous section, we have that in each case we have:

\begin{itemize}
\item Case 1: $v \in \E^4 + \sum_{j=1}^6 r_j \cdot \R_+$ where $r_1, \hdots,
r_6$ are the ranks of the following matroids \revisiontwo{which respectively correspond to the variables $x, y, z, w, \revisiontwo{q}, t$}:
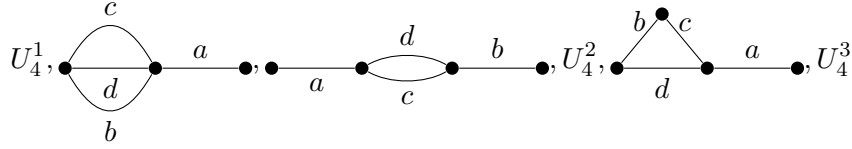
$$\revision{ U^1_4 },
\begin{tikzpicture}[baseline=0ex, scale=.6]
  \tikzstyle{vertex}=[circle,fill=black,minimum size=5pt,inner sep=0pt];
  \node[vertex] (a) at (0,0) {};
  \node[vertex] (b) at (2,0)  {};
  \node[vertex] (c) at (4,0) {};
  \draw (a) .. controls +(-60:40pt) and +(-120:40pt) .. node[below]{$b$}(b);
  \draw (a) .. controls +(+60:40pt) and +(+120:40pt) .. node[above]{$c$} (b);
  \draw (a) -- node[below] {$d$}(b);
  \draw (b) -- node[above] {$a$}(c);
\end{tikzpicture},
\begin{tikzpicture}[baseline=0ex, scale=.6]
  \tikzstyle{vertex}=[circle,fill=black,minimum size=5pt,inner sep=0pt];
  \node[vertex] (a) at (0,0) {};
  \node[vertex] (b) at (2,0)  {};
  \node[vertex] (c) at (4,0) {};
  \node[vertex] (d) at (-2,0) {};
  \draw (a) .. controls +(-30:20pt) and +(-150:20pt) .. node[below]{$c$}(b);
  \draw (a) .. controls +(+30:20pt) and +(+150:20pt) .. node[above]{$d$}(b);
  \draw (a) -- node[below] {$a$}(d);
  \draw (b) -- node[above] {$b$}(c);
\end{tikzpicture}, \revision{U_4^2}, 
\begin{tikzpicture}[baseline=0ex, scale=.6]
  \tikzstyle{vertex}=[circle,fill=black,minimum size=5pt,inner sep=0pt];
  \node[vertex] (a) at (0,0) {};
  \node[vertex] (b) at (2,0)  {};
  \node[vertex] (c) at (1,1.2) {};
  \node[vertex] (d) at (4,0) {};
  \draw (a) -- node[above] {$b$}(c);
  \draw (b) -- node[above] {$c$}(c);
  \draw (a) -- node[below] {$d$}(b);
  \draw (b) -- node[above] {$a$}(d);
\end{tikzpicture}, \revision{ U_4^3 }
$$
\item Case 2: Same as before but with the rank functions of the following
matroids \revisiontwo{which respectively correspond to the variables $x, y, z, w, \revisiontwo{q}, t$}:
    $$ \revision{U^1_4},
\begin{tikzpicture}[baseline=0ex, scale=.6]
  \tikzstyle{vertex}=[circle,fill=black,minimum size=5pt,inner sep=0pt];
  \node[vertex] (a) at (0,0) {};
  \node[vertex] (b) at (2,0)  {};
  \node[vertex] (c) at (4,0) {};
  \draw (a) .. controls +(-60:40pt) and +(-120:40pt) .. node[below]{$b$}(b);
  \draw (a) .. controls +(+60:40pt) and +(+120:40pt) .. node[above]{$c$} (b);
  \draw (a) -- node[below] {$d$}(b);
  \draw (b) -- node[above] {$a$}(c);
\end{tikzpicture},
\begin{tikzpicture}[baseline=0ex, scale=.6]
  \tikzstyle{vertex}=[circle,fill=black,minimum size=5pt,inner sep=0pt];
  \node[vertex] (a) at (0,0) {};
  \node[vertex] (b) at (2,0)  {};
  \node[vertex] (c) at (4,0) {};
  \node[vertex] (d) at (-2,0) {};
  \draw (a) .. controls +(-30:20pt) and +(-150:20pt) .. node[below]{$c$}(b);
  \draw (a) .. controls +(+30:20pt) and +(+150:20pt) .. node[above]{$d$}(b);
  \draw (a) -- node[below] {$a$}(d);
  \draw (b) -- node[above] {$b$}(c);
\end{tikzpicture},\revision{ U_4^2}, 
\begin{tikzpicture}[baseline=0ex, scale=.6]
  \tikzstyle{vertex}=[circle,fill=black,minimum size=5pt,inner sep=0pt];
  \node[vertex] (a) at (0,0) {};
  \node[vertex] (b) at (2,0)  {};
  \node[vertex] (c) at (1,1.2) {};
  \node[vertex] (d) at (4,0) {};
  \draw (a) -- node[above] {$a$}(c);
  \draw (b) -- node[above] {$c$}(c);
  \draw (a) -- node[below] {$d$}(b);
  \draw (b) -- node[above] {$b$}(d);
\end{tikzpicture},  \revision{U_4^3}
$$
\item Case 3a: Same as before but with the rank functions of the following
matroids \revisiontwo{which respectively correspond to the variables $x, y, z, w, \revisiontwo{q}, t$}:
$$ \revision{U^1_4},
\begin{tikzpicture}[baseline=0ex, scale=.6]
  \tikzstyle{vertex}=[circle,fill=black,minimum size=5pt,inner sep=0pt];
  \node[vertex] (a) at (0,0) {};
  \node[vertex] (b) at (2,0)  {};
  \node[vertex] (c) at (4,0) {};
  \draw (a) .. controls +(-60:40pt) and +(-120:40pt) .. node[below]{$b$}(b);
  \draw (a) .. controls +(+60:40pt) and +(+120:40pt) .. node[above]{$c$} (b);
  \draw (a) -- node[below] {$d$}(b);
  \draw (b) -- node[above] {$a$}(c);
\end{tikzpicture},
\begin{tikzpicture}[baseline=0ex, scale=.6]
  \tikzstyle{vertex}=[circle,fill=black,minimum size=5pt,inner sep=0pt];
  \node[vertex] (a) at (0,0) {};
  \node[vertex] (b) at (2,0)  {};
  \node[vertex] (c) at (1,1.2) {};
  \draw (a) .. controls +(-30:20pt) and +(-150:20pt) .. node[below]{$c$}(b);
  \draw (a) .. controls +(+30:20pt) and +(+150:20pt) .. node[above]{$d$}(b);
  \draw (a) -- node[above] {$a$}(c);
  \draw (b) -- node[above] {$b$}(c);
\end{tikzpicture}, \revision{U_4^2}, 
\begin{tikzpicture}[baseline=0ex, scale=.6]
  \tikzstyle{vertex}=[circle,fill=black,minimum size=5pt,inner sep=0pt];
  \node[vertex] (a) at (0,0) {};
  \node[vertex] (b) at (2,0)  {};
  \node[vertex] (c) at (1,1.2) {};
  \node[vertex] (d) at (4,0) {};
  \draw (a) -- node[above] {$a$}(c);
  \draw (b) -- node[above] {$b$}(c);
  \draw (a) -- node[below] {$d$}(b);
  \draw (b) -- node[above] {$c$}(d);
\end{tikzpicture},  \revision{U_4^3}
$$
\item Case 3b: Same as before but with the rank functions of the following
matroids \revisiontwo{which respectively correspond to the variables $x, y, z, w, \revisiontwo{q}, t$}:
$$ \revision{U^1_4},
\begin{tikzpicture}[baseline=0ex, scale=.6]
  \tikzstyle{vertex}=[circle,fill=black,minimum size=5pt,inner sep=0pt];
  \node[vertex] (a) at (0,0) {};
  \node[vertex] (b) at (2,0)  {};
  \node[vertex] (c) at (4,0) {};
  \draw (a) .. controls +(-60:40pt) and +(-120:40pt) .. node[below]{$b$}(b);
  \draw (a) .. controls +(+60:40pt) and +(+120:40pt) .. node[above]{$c$} (b);
  \draw (a) -- node[below] {$d$}(b);
  \draw (b) -- node[above] {$a$}(c);
\end{tikzpicture},
\begin{tikzpicture}[baseline=0ex, scale=.6]
  \tikzstyle{vertex}=[circle,fill=black,minimum size=5pt,inner sep=0pt];
  \node[vertex] (a) at (0,0) {};
  \node[vertex] (b) at (2,0)  {};
  \node[vertex] (c) at (1,1.2) {};
  \draw (a) .. controls +(-30:20pt) and +(-150:20pt) .. node[below]{$c$}(b);
  \draw (a) .. controls +(+30:20pt) and +(+150:20pt) .. node[above]{$d$}(b);
  \draw (a) -- node[above] {$a$}(c);
  \draw (b) -- node[above] {$b$}(c);
\end{tikzpicture}, \revision{U_4^2}, 
\begin{tikzpicture}[baseline=0ex, scale=.6]
  \tikzstyle{vertex}=[circle,fill=black,minimum size=5pt,inner sep=0pt];
  \node[vertex] (a) at (0,0) {};
  \node[vertex] (b) at (2,0)  {};
  \node[vertex] (c) at (1,1.2) {};
  \node[vertex] (d) at (4,0) {};
  \draw (a) -- node[above] {$a$}(c);
  \draw (b) -- node[above] {$b$}(c);
  \draw (a) -- node[below] {$c$}(b);
  \draw (b) -- node[above] {$d$}(d);
\end{tikzpicture},  \revision{U_4^3}
$$
\item Case 3c: Same as before but with the rank functions of the following
matroids \revisiontwo{which respectively correspond to the variables $x, y, \revisiontwo{q}, w, \revisiontwo{s}, t$}:
$$ \revision{U^1_4},
\begin{tikzpicture}[baseline=0ex, scale=.6]
  \tikzstyle{vertex}=[circle,fill=black,minimum size=5pt,inner sep=0pt];
  \node[vertex] (a) at (0,0) {};
  \node[vertex] (b) at (2,0)  {};
  \node[vertex] (c) at (4,0) {};
  \draw (a) .. controls +(-60:40pt) and +(-120:40pt) .. node[below]{$b$}(b);
  \draw (a) .. controls +(+60:40pt) and +(+120:40pt) .. node[above]{$c$} (b);
  \draw (a) -- node[below] {$d$}(b);
  \draw (b) -- node[above] {$a$}(c);
\end{tikzpicture},
\revision{
\begin{tikzpicture}[baseline=0ex, scale=.6]
  \tikzstyle{vertex}=[circle,fill=black,minimum size=5pt,inner sep=0pt];
  \node[vertex] (a) at (0,0) {};
  \node[vertex] (b) at (2,0)  {};
  \node[vertex] (c) at (1,1.2) {};
  \draw (a) .. controls +(-30:20pt) and +(-150:20pt) .. node[below]{$c$}(b);
  \draw (a) .. controls +(+30:20pt) and +(+150:20pt) .. node[above]{$d$}(b);
  \draw (a) -- node[above] {$a$}(c);
  \draw (b) -- node[above] {$b$}(c);
\end{tikzpicture},} \revision{U_4^2}, 
\revision{ \begin{tikzpicture}[baseline=0ex, scale=.6]
  \tikzstyle{vertex}=[circle,fill=black,minimum size=5pt,inner sep=0pt];
  \node[vertex] (a) at (0,0) {};
  \node[vertex] (b) at (2,0)  {};
  \node[vertex] (c) at (4,0) {};
  \node[vertex] (d) at (-2,0) {};
  \draw (a) .. controls +(-30:20pt) and +(-150:20pt) .. node[below]{$c$}(b);
  \draw (a) .. controls +(+30:20pt) and +(+150:20pt) .. node[above]{$d$}(b);
  \draw (a) -- node[below] {$a$}(d);
  \draw (b) -- node[above] {$b$}(c);
\end{tikzpicture}},  \revision{U_4^3}.
$$

\end{itemize}

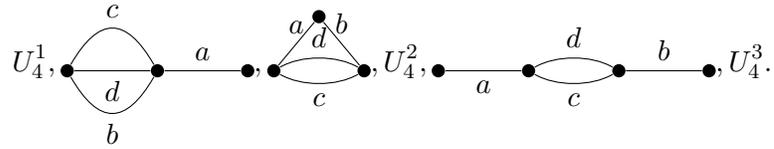
\end{document}